\documentclass[iop]{emulateapj}

\newcommand{\noprint}[1]{}

\newcommand{\kms}{\,\mathrm{km}\,\mathrm{s}^{-1}}

\newcommand{\yr}{\,\mathrm{yr}}
\newcommand{\kyr}{\,\mathrm{kyr}}
\newcommand{\Myr}{\,\mathrm{Myr}}

\newcommand{\Kv}{\,\mathrm{K}}

\newcommand{\GHz}{\,\mathrm{GHz}}
\newcommand{\mJy}{\,\mathrm{mJy}}

\newcommand{\kpc}{\,\mathrm{kpc}}
\newcommand{\pc}{\,\mathrm{pc}}
\newcommand{\invkpc}{\,\mathrm{kpc}^{-1}}

\newcommand{\erg}{\,\mathrm{erg}}
\newcommand{\ergs}{\,\mathrm{erg}\,\mathrm{s}^{-1}}
\newcommand{\cmq}{\,\mathrm{cm}^{-3}}

\newcommand{\Msun}{\,\mathrm{M}_{\sun}}

\newcommand{\kB}{k_\mathrm{B}}
\newcommand{\vs}{v_s}
\newcommand{\Ts}{T_s}
\newcommand{\ns}{n_s}

\newcommand{\vrw}{\left<v_{r,w}\right>}
\newcommand{\vout}{v_\mathrm{out}}
\newcommand{\vrad}{v_r}
\newcommand{\kmin}{k_\mathrm{min}}
\newcommand{\Pjet}{P_\mathrm{jet}}
\newcommand{\Pjetff}{P_\mathrm{jet,45}}

\newcommand{\pjet}{p_\mathrm{jet}}
\newcommand{\vjet}{v_\mathrm{jet}}
\newcommand{\rhojet}{\rho_\mathrm{jet}}

\newcommand{\Mwtot}{M_{w,\mathrm{tot}}}
\newcommand{\rhohot}{\rho_h}
\newcommand{\nhot}{n_h}
\newcommand{\Thot}{T_h}

\newcommand{\muhot}{\mu_\mathrm{m}}
\newcommand{\nwarmav}{\left<n_w\right>}
\newcommand{\Rcloud}{R_c}
\newcommand{\Rcmax}{R_\mathrm{c,max}}
\newcommand{\Rbulge}{R_\mathrm{bulge}}
\newcommand{\taujc}{\tau_\mathrm{jc}}
\newcommand{\rhocloud}{\rho_c}

\newcommand{\vabl}{v_\mathrm{abl}}
\newcommand{\pism}{p_\mathrm{ISM}}
\newcommand{\Tcrit}{T_\mathrm{crit}}
\newcommand{\rhocrit}{\rho_\mathrm{crit}}
\newcommand{\ncrit}{n_\mathrm{crit}}
\newcommand{\Ekinw}{E_{\mathrm{kin},w}}
\newcommand{\phiw}{\phi_w}

\newcommand{\fvol}{f_V}
\newcommand{\fgas}{f_\mathrm{gas}}
\newcommand{\few}{\mathrm{few}}

\newcommand{\Ledd}{L_\mathrm{edd}}
\newcommand{\Thomsonx}{\sigma_T}
\newcommand{\mprot}{m_p}

\newcommand{\MBH}{M_\mathrm{BH}}
\newcommand{\Mbulge}{M_\mathrm{\star}}

\newcommand{\mdotout}{\dot{M}_\mathrm{out}}

\newcommand{\fw}{f_w}
\newcommand{\etacrit}{\eta_\mathrm{crit}}
\newcommand{\etaobs}{\eta_\mathrm{obs}}
\newcommand{\vch}{v_\mathrm{ch}}
\newcommand{\vchentr}{v_\mathrm{ch,entr}}
\newcommand{\vchabl}{v_\mathrm{ch,abl}}
\newcommand{\mcdot}{\dot{m}_c}
\newcommand{\mcdotmax}{\dot{m}_{c\mathrm{,max}}}
\newcommand{\nch}{n_\mathrm{ch}}
\newcommand{\vexp}{v_\mathrm{exp}}

\newcommand{\fentr}{f_\mathrm{entr}}
\newcommand{\tacc}{t_\mathrm{acc}}
\newcommand{\Mdotjet}{\dot{M}_\mathrm{jet}}
\newcommand{\Mdotentr}{\dot{M}_\mathrm{entr}}
\newcommand{\Mdotabl}{\dot{M}_\mathrm{abl}}
\newcommand{\Mdottot}{\dot{M}_\mathrm{tot}}
\newcommand{\Mch}{\mathcal{M}_\mathrm{ch}}

\newcommand{\ud}{\mathrm{d}}
\newcommand{\td}[2]{\frac{d#1}{d#2}}
\newcommand{\erf}{\mathrm{erf}}

\newcommand{\HI}{H\,{\sc i}}

\newcommand{\OIII}{O\,{\sc iii}}

\newcommand{\NeII}{Ne\,{\sc ii}}

\newcommand{\eqref}[1]{(\ref{#1})}

\newcommand{\btr}{\blacktriangleright}

\newcommand{\msigma}{$M$--$\sigma$}

\newcommand{\pasa}{PASA}
\newcommand{\nar}{New Astron. Rev.}

\slugcomment{Accepted by the Astrophysical Journal on August 9, 2012.}

\shorttitle{Relativistic Jet Feedback}
\shortauthors{Wagner, Bicknell, \&{} Umemura}

\begin{document}


\title{Driving Outflows with Relativistic Jets and the Dependence of AGN Feedback Efficiency on ISM Inhomogeneity}


\author{A.~Y. Wagner\altaffilmark{1}, G.~V. Bicknell\altaffilmark{2}, M. Umemura\altaffilmark{1}}
\altaffiltext{1}{Center for Computational Sciences, University of Tsukuba, 1-1-1 Tennodai, Tsukuba, Ibaraki, 305-8577, Japan}
\altaffiltext{2}{Research School of Astronomy and Astrophysics, The Australian National University, ACT 2611, Australia}
\email{ayw@ccs.tsukuba.ac.jp}


\begin{abstract}
We examine the detailed physics of the feedback mechanism by relativistic AGN jets interacting with a two-phase fractal interstellar medium in the kpc-scale core of galaxies using 29 3D grid-based hydrodynamical simulations. The feedback efficiency, as measured by the amount of cloud-dispersal generated by the jet-ISM interactions, is sensitive to the maximum size of clouds in the fractal cloud distribution but not to their volume filling factor. Feedback ceases to be efficient for Eddington ratios $\Pjet/\Ledd\lesssim10^{-4}$, although systems with large cloud complexes $\gtrsim50\pc$ require jets of Eddington ratio in excess of $10^{-2}$ to disperse the clouds appreciably. Based on measurements of the bubble expansion rates in our simulations we argue that sub-grid AGN prescriptions resulting in negative feedback in cosmological simulations without a multi-phase treatment of the ISM are good approximations if the volume filling factor of warm phase material is less than 0.1 and the cloud complexes are smaller than $\sim25\pc$. We find that the acceleration of the dense embedded clouds is provided by the ram pressure of the high velocity flow through the porous channels of the warm phase, flow that has fully entrained the shocked hot-phase gas it has swept up, and is additionally mass-loaded by ablated cloud material. This mechanism transfers $10\%$ to $40\%$ of the jet energy to the cold and warm gas, accelerating it within a few 10 to 100 $\Myr$ to velocities that match those observed in a range of high and low redshift radio galaxies hosting powerful radio jets.\end{abstract}








\section{Introduction}
\label{s:intro}

The formation of galaxies is a non-linear, but to some degree self-regulatory process; the star-formation efficiencies of galaxies and the growth rate of the central supermassive black-holes (SMBH) are thought to be modified by feedback processes from active galactic nuclei (AGN) resulting in a tight correlation between SMBH mass and the bulge stellar velocity dispersion \citep[the \msigma{} relation,][]{ferrarese2000a,gebhardt2000a,tremaine2002a}.

It is unclear, however, which types of AGN activity are relevant in regulating bulge and SMBH growth. The \citet{silk1998a} model invokes an energy-driven quasar wind of low Eddington ratio, while the models by \citet{fabian1999a}, \citet{king2003a}, and \citet{murray2005a} consider opacity-regulated momentum-driven outflows requiring Eddington ratios of a few percent. Another possibility is that the radiation field in the bulge of galaxies controls the accretion rate of matter into the central regions \citep{umemura2001a, kawakatu2002a}. Cosmological SPH and semi-analytic models routinely include feedback by powerful radio jets or quasar winds, albeit, of necessity, using highly simplified models for the feedback. Observationally, ionization diagnostics may not conclusively distinguish the contributions of radiatively driven feedback and feedback driven by jet-ISM interactions \citep{holt2009b, hayashi2012a}, although in some cases jet-ISM interactions are strongly favoured \citep{dopita1997a,nesvadba2010a}. Several studies find statistical correlations between AGN activity, outflows, and the suppression of star-formation \citep[e.g.][]{schawinski2007a,farrah2012a}, but the connection between AGN jets and star-formation remains ambiguous \citep{dicken2012a, hayashi2012a}.

In cosmological SPH simulations \citep[e.g.][]{okamoto2008b,di-matteo2008a,schaye2010a}, grid-based simulations \citep[e.g.][]{springel2011a, dubois2012a}, and semi-analytic models \citep[e.g.][]{croton2006a,fanidakis2012a} AGN feedback is found to be a necessary ingredient in order to reproduce the observed galaxy luminosity function and its evolution with redshift, but the relevant range of powers varies between models. Cosmological SPH simulations require energy injection rates described by Eddington ratios $\eta\gtrsim10^{-2}$ while some semi-analytic models find that low-powered injection of energy with Eddington ratios of $\eta\gtrsim10^{-5}$ is sufficient. In both methods there exist a variety of \lq\lq{}subgrid\rq\rq{} prescriptions to deposit energy yielding different results. Neither method resolves or treats the galaxy-scale physics of the interaction of the outflows and interstellar medium (ISM) adequately, and one of our aims is to provide a a robust description of sub-grid feedback physics that can be used in future semi-analytic and cosmological models.

Feedback involving mechanical energy input by an AGN jet, often termed \lq\lq{}radio-mode\rq\rq{} feedback, has been identified as a key mechanism to heat the IGM of the cluster \citep[e.g.][]{binney1995a,soker2001a} and prevent a runaway build-up of galaxy mass through further accretion of cooling gas \citep[see][and references therein]{best2006a}. Well-studied nearby examples include the Hydra A \citep{wise2007b}, Perseus A \citep{fabian2006a}, and M87 in the Virgo cluster \citep{million2010a}, and the phenomenon is well reproduced in cluster-scale grid-based hydrodynamic simulations by \citet{gaspari2012a}, \citet{dubois2010a, dubois2011a} and \citet{teyssier2011a}. 

Galaxy-scale jet-regulated star-formation (\lq\lq{}positive\rq\rq{} feedback) may be very relevant at higher redshifts in gas-rich galaxies and proto-galactic environments \citep{de-young1989a, bicknell2000a, reuland2003a, klamer2004a, miley2006a, villar-martin2007a, venemans2007a, miley2008a, hayashi2012a}. The case for jet-induced star-formation in disc galaxies was made in numerical work as early as \citet{woodward1976a} and in more recent simulations by \citet{fragile2005a} and \citet{gaibler2011a}.

In some nearby and high-redshift radio galaxies (HzRG) neutral and line-emitting gas is seen outflowing at several $100\kms$ to several $1000\kms$ \citep{gelderman1994a,tadhunter2001a,odea2002a,emonts2005a,holt2008a,holt2011a,morganti2005a,morganti2007a,morganti2010a,nesvadba2006a,nesvadba2007a,nesvadba2008a,nesvadba2010a,lehnert2011a,dasyra2012a,guillard2012a,torresi2012a}. The alignment of the jet with outflowing gas \citep{pentericci2001a,privon2008a}, and matching energetics \citep{nesvadba2006a,nesvadba2007a} suggest that the outflows are driven by the transfer of energy and momentum from the jet to to the dense ISM. This hypothesis is supported by our previous 3D hydrodynamical simulations of AGN-jet driven outflows \citep[][WB11 henceforth]{wagner2011a}.

The question of how a collimated jet may impart energy and momentum isotropically, e.g., to affect the entire volume defining the bulge of a galaxy, is frequently mentioned \citep{de-young2010a, ostriker2010a}. A related problem is the momentum budget associated with the dispersion or expulsion of clouds in a galaxy. An important feature of AGN jets is that the jet is extremely light and that jet and cocoon are highly overpressured (underexpanded) with respect to the ambient environment \citep{begelman1989a}. Simulations by \citet{saxton2005a}, \citet{sutherland2007a}, \citet{gaibler2009a}, and WB11 of AGN jets show how the light, overpressured jet inflates a cocoon that drives a quasi-spherical energy bubble into the ISM. These simulations also showed that isotropization of the injected energy is even more effective when the jet encounters inhomogeneities because, by virtue of its lightness -- the jet particle density is typically 6 to 8 orders of magnitude lower than that of ISM clouds -- the jet is strongly deflected by the inhomogeneities. Additional effects, e.g., jet instabilities and jet precession increase the isotropy of energy deposition, but are not essential.

In previous work (WB11) we used grid-based hydrodynamic simulations to model jet-ISM interactions and quantified the feedback efficiency provided by relativistic AGN jets in the core of young, gas-rich radio galaxies. The simulated galaxies typically represent either Compact Steep Spectrum (CSS) or Gigahertz Peaked Spectrum (GPS) sources \citep{bicknell1997a}, which in our view are a class of objects experiencing an early phase of powerful jet-mediated feedback. In these objects, radio source expansion is impeded by the dense multi-phase environment of the galaxy core in the early phase of their evolution. We concluded that AGN jet feedback in these systems is effective in all galaxies for jets with powers $10^{43}$ -- $10^{46}\ergs$ if the ratio of jet power to Eddington luminosity $\eta\gtrsim10^{-4}$.

A unique feature of these simulations is the treatment of the galaxy ISM with a two-point fractal, single point log-normal warm-phase distribution (clouds) embedded in a hot atmosphere. We determined feedback efficiencies as a function of some of the parameters describing this distribution, e.g. the density and filling factor of the warm gas. The ISM properties in HzRG are uncertain; while large reservoirs of molecular gas and HI are known to exist, the volume filling factors of the cold and warm gas and the typical sizes of clouds are not known. WB11 restricted themselves to volume filling factors of the clouds of 0.42 and 0.13, which are probably at the higher end of the range of \lq\lq{}typical\rq\rq{} values. Furthermore, we did not investigate the dependence on maximum cloud sizes. 

With the 15 new simulations presented in this paper, we have now substantially extended this parameter space study to lower filling factors and a variety of maximum cloud sizes in the fractal distribution. We also examine the acceleration mechanism in more detail, providing an explanation for the high mechanical advantage observed by WB11. We describe our methods of computation and parameter space next in \S\ref{s:code} and \S\ref{s:params}, and compare the relevant timescales in the problem in \S\ref{s:timescales}. We present our main results in detail in \S\ref{s:results}. In \S\ref{s:disc}, we compare our simulation results with data from a sample of radio galaxies with observed outflows. We also discuss other feedback criteria and the review the difficulties in modelling cloud ablation. We conclude with a summary of the paper in \S\ref{s:concl}.

\section{Equations and code}\label{s:code}

The system of equations describing the relativistic jet plasma, hot atmosphere, and warm clouds in the one fluid approximation is \citep{landau1987a}:
\begin{eqnarray}
  \frac{\partial D}{\partial t} + \frac{\partial Du^i}{\partial x^i} = 0;  \quad  && D = \Gamma \rho \nonumber\;; \\
  \frac{\partial F^i}{\partial t} + \frac{\partial F^i u^j}{\partial x^j} + \frac{\partial p}{\partial x^i} = 0; \quad  && F^i = \rho w \Gamma^2 u^i/c^2\;; \label{e:rel}\\
  \frac{\partial E}{\partial t} + \frac{\partial F^i c^2 }{\partial x^i}  = -\rho^2\Lambda(T); \quad  && E = \rho w \Gamma^2 - p. \nonumber
\end{eqnarray}

The conserved quantities, $D$, $F^i$, and $E$ are the laboratory frame fluid density, components of the momentum density, total energy density (including the rest mass energy density). The variables $p$, $T$, $\Lambda$, and $u^i$ are pressure, temperature, cooling rate, and the components of the three velocity, respectively. The bulk Lorentz factor is $\Gamma = \left( 1 - u_iu^i/c^2 \right)^{-1/2}$. The proper rest frame density is $\rho$ and $w= c^2 + p\gamma/\rho(\gamma - 1)$ the proper rest frame specific enthalpy for an ideal polytropic equation of state, with index $\gamma$.

We integrate these equations using the publicly available, open-source Eulerian Godunov-type code FLASH \citep{fryxell2000a} version 3.2 and its relativistic hydrodynamics module \citep{mignone2005b} to which we have added code to incorporate radiative cooling and code to advance advected scalars in the relativistic hydrodynamic solver.

We exploit the adaptive mesh capabilities of FLASH, utilizing up to seven levels of grid refinement in a cubical simulation domain of $1\kpc^{3}$ in physical dimensions, consisting of $1024^3$ cells at a maximum spatial resolution $1\pc$. This is twice the resolution of the simulations by WB11 and is necessary in order to capture the fractal outlines of clouds for small filling factors and cloud sizes. The jet inlet and initial jet width is $20\pc$ and is resolved with at least 10 cells. The formation of clean diamond shocks indicates that the jet stream is sufficiently well resolved. Note that a restricted one parameter scaling of physical dimensions is possible \citep{sutherland2007a}. 

Tracer variables distinguish jet material and warm phase gas from each other and from the hot phase background. We include non-equilibrium, optically thin atomic cooling for $T>10^4\Kv$ \citep{sutherland1993a} and updated solar abundances \citep{asplund2005a}, for which the mean mass per particle, $\muhot=0.6165$. Thermal conduction, photo-evaporation, self-gravity, and magnetic fields are not included. 

We do not include a static gravitational potential for the SMBH or the bulge because our simulations span a spatial range from $1\pc$ -- $1\kpc$, within which neither the gravitational force due to the SMBH, nor that due to the bulge are dynamically important over the timescales considered here. For typical SMBH and bulge masses in evolving massive galaxies, the SMBH sphere of influence extends to a radial distance of order $10\pc$, covering only a few tens of cells in our simulation domain around the base of the jet. This volume is quickly filled with light jet plasma, which is practically unaffected by gravity. On the $\kpc$ scale, the density and pressure profiles of the hydrostatic environment in a massive gas-rich spheroidal proto-galaxy are fairly flat under the gravitational influence of the bulge \citep[e.g.][]{capelo2010a}, and we adopt a uniform hot phase distribution characterized by a temperature of $\Thot=10^7\Kv$ and a density of either $\nhot=0.1$ or $\nhot=1.0$. The gravitational force due to the bulge may be neglected from timescale arguments, described in \S\ref{s:timescales}.

We ran our simulations on the National Computational Infrastructure National Facility (NCI NF) Oracle/Sun Constellation Cluster, a high-density integrated system of 1492 nodes of Sun X6275 blades, each containing two quad-core $2.93\GHz$ Intel Nehalem cpus, and four independent SUN DS648 Infiniband switches.\footnote{For details of the system specifications see http://nf.nci.org.au/facilities/vayu/hardware.php.} We typically used 256 to 1024 cpus with 3GB of memory per core to complete one simulation within two weeks.

\section{Model parameters, initial conditions, and boundary conditions}\label{s:params}

A crucial ingredient in the simulations described in \S\ref{s:params} is the two-phase ISM, which consists of a warm ($T \sim 10^4 \Kv$) phase and a hot ($T\sim 10^7 \Kv$) phase. In particular, we are concerned with the effect of the jet plasma on the state and dynamics of the warm phase material. We have, therefore, extended our studies of parameters related to the warm-phase and identified the correct physical mechanism that leads to the acceleration of the clouds.

The warm phase ISM density is initialized from a cube of random numbers that simultaneously satisfies single-point log-normal statistics and two-point fractal statistics. Let $P(\rho)$ be the log-normal probability density function of the random variable $\rho$, representing density:
\begin{equation}
P(\rho) = \frac{1}{s\sqrt{2\pi}\rho}\exp\left(\frac{-(\ln\rho - m)^2}{2s^2}\right)\;,\label{e:logn}
\end{equation}
where
\begin{equation}
m = \ln\frac{\mu^2}{\sqrt{\sigma^2 + \mu^2}} \;, \quad s = \sqrt{\ln\left(\frac{\sigma^2}{\mu^2} + 1\right)} \;, 
\end{equation}
and $\mu$ and $\sigma^2$ are the mean and variance of the log-normal distribution.

Let $F(\mathbf{k})$ be the Fourier transform of the spatial density distribution, $\rho(\mathbf{r})$, with $\mathbf{k}$ and $\mathbf{r}$ as wave vector and position vector, respectively. The two-point fractal property is characterized in Fourier space by a power spectrum, $D(k)$, in wave number, $k$, that obeys a power-law with index $-5/3$ for a Kolmogorov-type spectrum
\begin{equation}
D(k) = \int k^2 F(\mathbf{k}) F^{*}(\mathbf{k}) \ud\Omega \propto k^{-5/3}\;,\label{e:frac}
\end{equation}
where the integral of the spectral density, $F(\mathbf{k}) F^{*}(\mathbf{k})$, is over all solid angle, $\Omega$.

A cube of random numbers that simultaneously satisfies Eqn.~\eqref{e:logn} and Eqn.~\eqref{e:frac} is generated by the method outlined in \citet{lewis2002a}. First, a cube with cell values from a Gaussian distribution with mean $m$ and standard deviation $s$ is Fourier-transformed and apodized by a Kolmogorov power law spectrum in wavenumber with index $-5/3$ and minimum sampling wavenumber $\kmin$. The minimum sampling wavenumber is, effectively, the average number of clouds per dimension divided by 2, and it determines the scale of the largest fractal structures in the cube relative to the size of the cube. For example, if $\kmin=20$ for a cube mapped to a domain of extent $1\kpc$, then $\kmin=20\invkpc$ and the largest structures (clouds) would have extents of $\Rcmax=1/(2\kmin)=25\pc$. The cube is then transformed back into real space and exponentiated. Because the last step alters the power-law structure in Fourier space, the cube is iteratively transformed between Fourier space and real space until successive corrections produce a power-law convergence within 1\%.

To place the fractal cube into the simulation domain it is apodized (in real space) by a spherically symmetric mean density profile which in the simulations presented here is flat with mean warm phase density $\nwarmav$. The porosity of the warm phase arises by imposing an upper temperature cutoff for the existence of clouds at $T_\mathrm{crit}=3\times10^4\Kv$, beyond which clouds are deemed thermally unstable. No lower temperature limit is enforced, and temperatures in the core of clouds may initially be less than $100\Kv$. The upper temperature cutoff corresponds directly to a lower density cutoff, $\rhocrit=\muhot p/(k \Tcrit)$, if the pressure, $p$, is defined. Here, $\muhot$ is the mean mass per particle of the hot phase. In our simulations the clouds are in pressure equilibrium with the surrounding hot phase, thus $\rhocrit=\muhot \nhot \Thot/\Tcrit$, where $\nhot$ and $\Thot$ are the hot phase number density and temperature, respectively. The filling factor of the warm phase, within the hemispherical region of radius $0.5\kpc$, in which it is distributed, is: 
\begin{eqnarray}
  \fvol&=&\int_{\rhocrit}^\infty P(\rho) \ud \rho \nonumber\\
  &=&\frac{1}{2}\left[1 + \erf\left(\frac{\ln\left\{(\rhocrit/\mu)\sqrt{\sigma^2/\mu^2 + 1}\right\}}{\sqrt{2\ln\left(\sigma^2/\mu^2 + 1\right)}} \right)\right]
\end{eqnarray}
The original fractal cube was constructed with $\mu=1$ and $\sigma=5$, and after apodization with a spatially uniform mean warm phase density distribution, the single point density distribution remains lognormal, but with a mean $\mu=\nwarmav$. For an isothermal hot phase distribution, whose temperature in this work is fixed at $\Thot=10^7$, $\rhocrit/\mu = (\Thot/\Tcrit) (\nhot/\nwarmav)$ is constant everywhere. The filling factor is, therefore, directly defined by the ratio of hot phase density and mean warm phase density, $\nhot/\nwarmav$. 

The clouds embedded in the hot phase remain static unless impacted by the jet or jet-blown bubble. The gas temperatures in most cells containing warm phase material are initially below $10^4\Kv$ and are not subject to radiative cooling. For a more detailed description of the method to generate the fractal cube and a discussion of the choice of statistical parameters for the lognormal probability distribution and wavenumber power law index we refer the reader to the manuscript by \citet{lewis2002a} and the relevant sections and appendixes in \citet{sutherland2007a}.

The general setup, initial conditions, and boundary conditions used here are identical to those of WB11. WB11 performed 14 simulations of AGN jets with powers in the range $43<\log(\Pjet/\ergs)<46$. The choice of warm phase filling factors, $\fvol$, of 0.42 and 0.13, was relatively high and the maximum cloud size fixed at $\Rcmax\sim25\pc$ ($\kmin=20\invkpc$). 

In the 15 new simulations presented here, we explore new regions of parameter space with filling factors, $\fvol$, of 0.052 and 0.027, corresponding to average warm phase densities of $150\cmq$ and $100\cmq$, and $\kmin$ of $40\invkpc$ and $10\invkpc$, corresponding to maximum cloud sizes of $\Rcmax=12.5\pc$ and $\Rcmax\sim 50.0\pc$. The range in jet power and other parameters defining the jet plasma remain the same. These jets typically have a density contrast of $10^{-4}$ with respect to the ambient hot phase and $10^{-7}$ with respect to the embedded clouds. The pressure contrast between the jet and the hot phase ISM is typically $10^2$ -- $10^3$. AGN jets are extremely light, underexpanded (overpressured) jets. 

The complete list of 29 simulations, including those from WB11 are given in Table~\ref{t:runs}. New runs are marked with \lq\lq$\btr$\rq\rq.

\section{Timescales}\label{s:timescales}

It is instructive to compare the timescales and associated lengthscales present in this problem. The definition and values of the relevant timescales are listed in Table~\ref{t:timescales}. 

An unimpeded jet typically requires of order $100\kyr$ to cross a domain of $1\kpc$, while a jet propagating through a clumpy ISM is confined anywhere between $20\kyr$ and $1\Myr$, depending on jet power and average cloud density.

In comparison to the confinement time, the bulge free-fall time is at least two orders of magnitude larger, justifying our neglect of gravity in the simulations. A closely related timescale, the buoyancy timescale for a jet blown bubble is also only important on timescales much longer than the simulation time and on spatial scales much larger than $1\kpc$ \citep{bruggen2002a}. 

The cooling time in the hot ISM is also longer than the simulation time, but the cooling time in the clouds is short enough to affect the computational hydrodynamic timestep and the cooling length is not resolved in our simulations. The sound crossing time inside clouds is much longer than that in the inter-cloud medium, and also much longer than the jet confinement time, and we use this property to slightly underpressure the clouds (by $\sim2\%$) to keep the cloud interfaces sharp and static. 

\citet{wagner2011a} compared the cloud collapse and cloud ablation timescales and concluded that, while clouds engulfed by the jet-blown bubble experienced an external pressure enhancement that reduced the critical Bonnor-Ebert mass sufficiently to formally induce collapse, the comparatively short ablation times may destroy clouds before stars can form. Cloud ablation is facilitated by the Kelvin-Helmholtz instability, which grows over timescales comparable to or shorter than the ablation timescale. The cloud crushing time is long in comparison. These timescales are included here for completeness.

\begin{deluxetable*}{lcc}
\tablecaption{Timescales in Jet-ISM interactions}
\tablehead{
\colhead{Timescale} & \colhead{Definition} & \colhead{Typical values}
}
\startdata
Unimpeded jet crossing time\tablenotemark{(a)}                    & $\frac{L}{\vjet}\left(1+\frac{1}{\Gamma}\sqrt{\frac{\chi}{(1+\chi)\zeta}}\right)$ & $6\kyr$ -- $300\kyr$ \\ 
Jet confinement time\tablenotemark{(b)}                           & \textemdash                                                                       & $20\kyr$ -- $1\Myr$  \\ 
Bulge free-fall time\tablenotemark{(c)}                           & $\sqrt{1/G\rhohot}$                                                               & $120\Myr$            \\ 
Buoyancy timescale\tablenotemark{(d)}                             & $0.5L/\sqrt{2gV/SC}$                                                              & $45\Myr$             \\ 
Cooling time in hot ISM                                           & $\kB \Thot/\nhot\Lambda(\Thot)$                                                   & $8.6\Myr$            \\ 
Cooling time in clouds at critical temperature\tablenotemark{(e)} & $\kB \Tcrit/\ncrit\Lambda(\Tcrit)$                                                & $15.5\yr$            \\ 
Cooling time in shocked clouds\tablenotemark{(f)}                 & $\kB \Ts/\ns\Lambda(\Ts)$                                                         & $14.6\yr$            \\ 
Cloud sound crossing time\tablenotemark{(g)}                      & $2\Rcloud/c_c$                                                                    & $6.6\Myr$            \\ 
Inter-cloud sound crossing time\tablenotemark{(h)}                & $d_{ch}/c_h$                                                                      & $66\kyr$             \\ 
Cloud Kelvin-Helmholtz growth time \tablenotemark{(i)}            & $(\Rcloud/\vch)(\nwarmav+\nch)/\sqrt{\nwarmav\nch}$                               & $24\kyr$             \\ 
Cloud crushing time \tablenotemark{(j)}                           & $\Rcloud/\vs\approx\nwarmav\Rcloud/\nch\vch$                                      & $2.4\Myr$            \\ 
Cloud collapse time\tablenotemark{(k)}                            & $\sqrt{1/G\rhocloud}$                                                             & $3.8\Myr$            \\ 
Cloud ablation time\tablenotemark{(l)}                            & $2\Rcloud/\vabl$                                                                  & $40\kyr$
\enddata
\tablenotetext{(a)}{Time required for the jet head to cross the $L=1\kpc$ domain if no clouds impeded its progress \citep[see e.g.][]{safouris2008a}. The variables $\Gamma$ and $\zeta$ denote the lorentz factor and the ratio of jet density to ambient gas density, and $\chi=(\gamma-1)\rhojet c^2/\gamma \pjet$ is the proper density parameter. The lower and upper values correspond to the cases for which $\Pjet=10^{46}\ergs$, $\nhot=0.1\cmq$ and $\Pjet=10^{43}\ergs$, $\nhot=1.0\cmq$, respectively.}
\tablenotetext{(b)}{Time required for the jet to cross the $L=1\kpc$ domain in the presence of clouds impeding its progress. This is effectively equivalent to the duration of the simulation.}
\tablenotetext{(c)}{$\rhohot=1.0\muhot$.}
\tablenotetext{(d)}{$V$, $S$, $C$, and $g$ are volume and cross-section of the buoyant bubble, the drag coefficient, and the gravitational acceleration, respectively \citep{birzan2004a}. Here we choose $V/S=0.5\kpc$ and $C=0.75$ \citep{churazov2001a}}
\tablenotetext{(e)}{$\ncrit$ is the critical number density corresponding to $\rhocrit$.}
\tablenotetext{(f)}{$\Ts$ and $\ns$ are the postshock temperature and particle number density, respectively, of the shock propagating into the cloud. A shock speed of $100\kms$ and preshock conditions of $n=\ncrit$ and $T=\Tcrit$ were assumed.} 
\tablenotetext{(g)}{$c_c$ is the average sound speed in a cloud with average temperature $1000\Kv$.}
\tablenotetext{(h)}{$c_h$ is the sound speed of the hot phase and $d_{ch}\sim2\Rcloud$ is the inter-cloud distance.}
\tablenotetext{(i)}{The growth timescale of the Kelvin-Helmholtz instability at the interface between a cloud and its ambient hot flow. $\vch\sim10^5\kms$ and $\nch\sim0.1\cmq$ are the velocity and particle number density of the flow through the inter-cloud channels. This is the upper limit corresponding to the lowest excited mode.}
\tablenotetext{(j)}{$\vs$ is the speed of the shock propagating into the cloud.}
\tablenotetext{(k)}{This is equivalent to the cloud free-fall time.}
\tablenotetext{(l)}{$\vabl\sim600\kms$ is the ablation speed. It corresponds to the channel speed observed in our simulations.}
\label{t:timescales}
\end{deluxetable*}

\section{Results}\label{s:results}

\begin{deluxetable*}{clcccccccc}
\tablecaption{Simulation parameters\label{t:runs}}
\tablewidth{\textwidth}
\tablehead{
\colhead{New} & \colhead{Simulation} & \colhead{$\log \, \Pjet$\tablenotemark{(a)}} & 
\colhead{$\nhot$\tablenotemark{(b)}} & \colhead{$\pism/k$\tablenotemark{(c)}} & 
\colhead{$\nwarmav$\tablenotemark{(d)}} & \colhead{§$\fvol$\tablenotemark{(e)}} & 
\colhead{$\kmin$\tablenotemark{(f)}} & \colhead{$\Rcmax$\tablenotemark{(g)}} & \colhead{$\Mwtot$\tablenotemark{(h)}}\\
\colhead{} & \colhead{} & \colhead{$(\erg)$} & \colhead{$(\cmq)$} & \colhead{$(\cmq\Kv)$} &
\colhead{$(\cmq)$} & \colhead{} & \colhead{$(\invkpc)$} & \colhead{$(\pc)$} & \colhead{$(10^{9}\Msun)$}
}
\startdata
      & A\dotfill                               & 45 & 0.1 & $10^6$ & \nodata& \nodata & \nodata & \nodata & \nodata        \\
      & B\dotfill                               & 46 & 1.0 & $10^7$ & 1000 & 0.42\phn  & 20      & 25.0    & 16\phd\phn\phn \\
      & B$^\prime$\dotfill                      & 46 & 1.0 & $10^7$ &  300 & 0.13\phn  & 20      & 25.0    & \phn3.2\phn    \\
$\btr$& B$^{\prime\prime}$\dotfill              & 46 & 1.0 & $10^7$ &  150 & 0.052     & 20      & 25.0    & \phn0.29       \\
$\btr$& B$^{\prime\prime\prime}$\dotfill        & 46 & 1.0 & $10^7$ &  100 & 0.027     & 20      & 25.0    & \phn0.15       \\
      & C\dotfill                               & 46 & 0.1 & $10^6$ &  100 & 0.42\phn  & 20      & 25.0    & \phn1.6\phn    \\
      & C$^\prime$\dotfill                      & 46 & 0.1 & $10^6$ &   30 & 0.13\phn  & 20      & 25.0    & \phn0.32       \\
      & D\dotfill                               & 45 & 1.0 & $10^7$ & 1000 & 0.42\phn  & 20      & 25.0    & 16\phd\phn\phn \\
$\btr$& D${\prime}_{10}$\dotfill               & 45 & 1.0 & $10^7$ &  300 & 0.13\phn  & 10      & 50.0    & \phn3.2\phn    \\
      & D$^\prime$\dotfill                      & 45 & 1.0 & $10^7$ &  300 & 0.13\phn  & 20      & 25.0    & \phn3.2\phn    \\
$\btr$& D$^{\prime\prime}_{10}$\dotfill         & 45 & 1.0 & $10^7$ &  150 & 0.052     & 10      & 50.0    & \phn0.29       \\
$\btr$& D$^{\prime\prime}$\dotfill              & 45 & 1.0 & $10^7$ &  150 & 0.052     & 20      & 25.0    & \phn0.29       \\
$\btr$& D$^{\prime\prime}_{40}$\dotfill         & 45 & 1.0 & $10^7$ &  150 & 0.052     & 40      & 12.5    & \phn0.29       \\
$\btr$& D$^{\prime\prime\prime}_{10}$\dotfill   & 45 & 1.0 & $10^7$ &  100 & 0.027     & 10      & 50.0    & \phn0.15       \\
$\btr$& D$^{\prime\prime\prime}$\dotfill        & 45 & 1.0 & $10^7$ &  100 & 0.027     & 20      & 25.0    & \phn0.15       \\
      & E\dotfill                               & 45 & 0.1 & $10^6$ &  100 & 0.42\phn  & 20      & 25.0    & \phn1.6\phn    \\
      & E$^\prime$\dotfill                      & 45 & 0.1 & $10^6$ &   30 & 0.13\phn  & 20      & 25.0    & \phn0.32       \\
      & F\dotfill                               & 44 & 0.1 & $10^6$ &  100 & 0.42\phn  & 20      & 25.0    & \phn1.6\phn    \\
      & F$^\prime$\dotfill                      & 44 & 0.1 & $10^6$ &   30 & 0.13\phn  & 20      & 25.0    & \phn0.32       \\
      & G\dotfill                               & 44 & 1.0 & $10^7$ & 1000 & 0.42\phn  & 20      & 25.0    & 16\phd\phn\phn \\
      & G$^\prime$\dotfill                      & 44 & 1.0 & $10^7$ &  300 & 0.13\phn  & 20      & 25.0    & \phn3.2\phn    \\
$\btr$& G$^{\prime\prime}_{10}$\dotfill         & 44 & 1.0 & $10^7$ &  150 & 0.052     & 10      & 50.0    & \phn0.29       \\
$\btr$& G$^{\prime\prime}$\dotfill              & 44 & 1.0 & $10^7$ &  150 & 0.052     & 20      & 25.0    & \phn0.29       \\
$\btr$& G$^{\prime\prime}_{40}$\dotfill         & 44 & 1.0 & $10^7$ &  150 & 0.052     & 40      & 12.5    & \phn0.29       \\
$\btr$& G$^{\prime\prime\prime}$\dotfill        & 44 & 1.0 & $10^7$ &  100 & 0.027     & 20      & 25.0    & \phn0.15       \\
      & H\dotfill                               & 43 & 0.1 & $10^6$ &  100 & 0.42\phn  & 20      & 25.0    & \phn1.6\phn    \\
$\btr$& H$^{\prime}$\dotfill                    & 43 & 0.1 & $10^6$ &   30 & 0.13\phn  & 20      & 25.0    & \phn1.6\phn    \\
$\btr$& I$^{\prime\prime}$\dotfill              & 43 & 1.0 & $10^7$ &  150 & 0.052     & 20      & 25.0    & \phn0.29       \\
$\btr$&I$^{\prime\prime\prime}$\dotfill         & 43 & 1.0 & $10^7$ &  100 & 0.027     & 20      & 25.0    & \phn0.15       
\enddata
\tablecomments{Runs with run labels containing the same letter are runs with the same jet power, $\Pjet$, and hot phase density, $\nhot$. Runs labeled with single, double, or triple primed (\lq\lq$\,^\prime\,$\rq\rq) letters denote lower filling factor counterparts to runs with less number of primes. All runs, other than those whose run label contains the value of $\kmin$ in the subscript, were performed with $\kmin=20\invkpc$.}
\tablenotetext{(a)}{Jet power.}
\tablenotetext{(b)}{Density of hot phase.}
\tablenotetext{(c)}{$p/k$ of both hot and warm phases.}
\tablenotetext{(d)}{Average density of warm phase.}
\tablenotetext{(e)}{Volume filling factor of warm phase.}
\tablenotetext{(f)}{Minimum sampling wave number.}
\tablenotetext{(g)}{Maximum cloud size.}
\tablenotetext{(h)}{Total mass in warm phase.}
\end{deluxetable*}

\subsection{Velocities of accelerated clouds and feedback efficiency}\label{s:sims}

\begin{figure*}
  \figurenum{1}
  \begin{center}
    \includegraphics[width=1.0\textwidth]{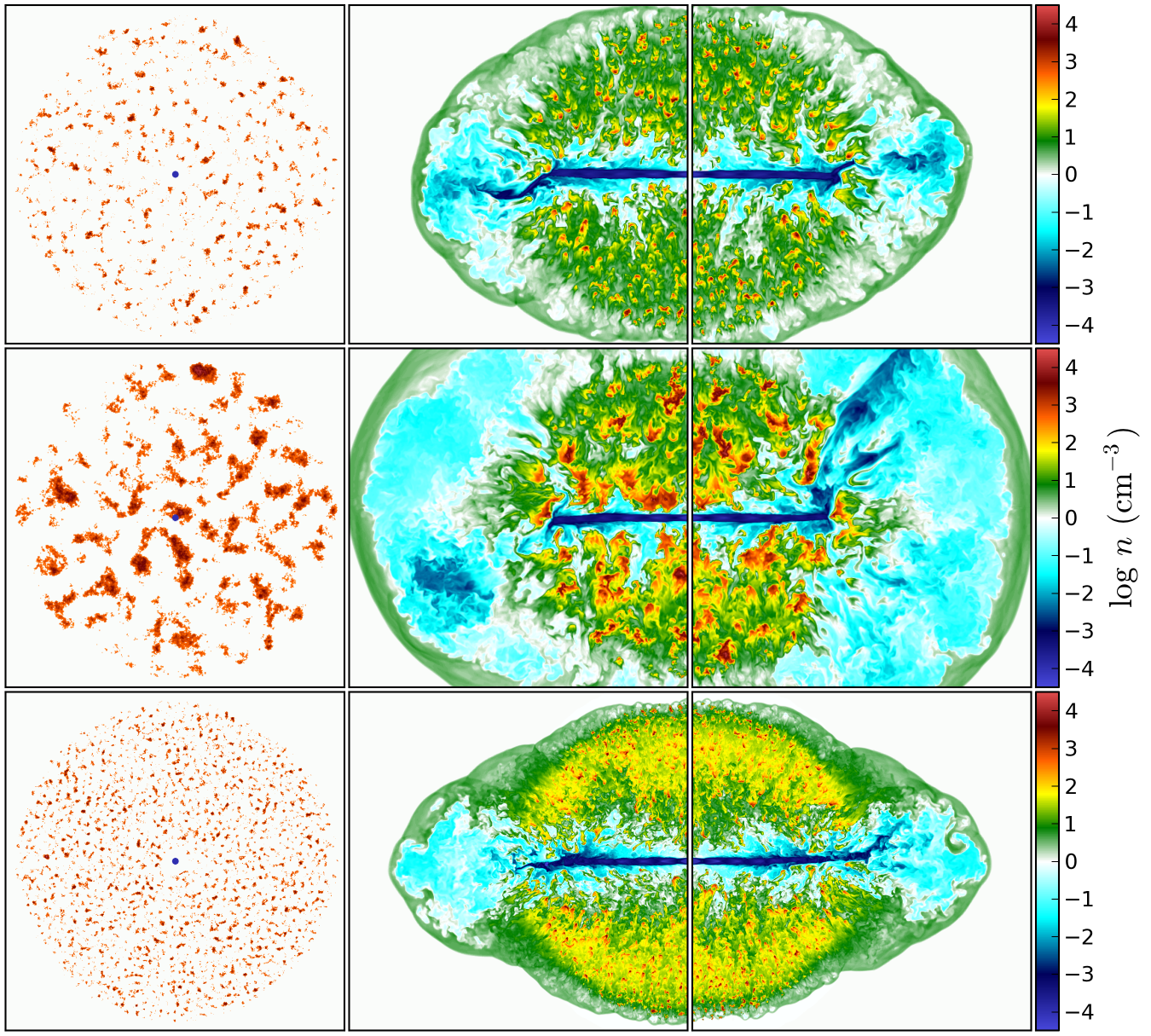}\\%
  \end{center}
  \caption{Logarithmic density maps (in units of $\cmq$) of selected new simulations. The domain extents in each panel are $1\kpc\times1\kpc$. The left column of panels show a face on view of initial the warm gas distribution. The center and right columns of panels show midplane slices at an advanced stage of the simulations for $z=0$ (reflected about $x=0$) and $y=0$, respectively. \textit{Top row}: Run D$^{\prime\prime\prime}$, a very low filling factor run ($\fvol=0.027$); \textit{Middle row}: Run D$^\prime_{10}$, maximum cloud sizes of $\Rcmax=50\pc$; \textit{Bottom row}: Run D$^{\prime\prime}_{40}$, maximum cloud sizes of $\Rcmax=10\pc$. See the electronic edition of the Journal for a color version of this figure.}
  \label{f:dens}
\end{figure*}

\begin{figure*}
  \figurenum{2}
  \includegraphics[width=1.0\textwidth]{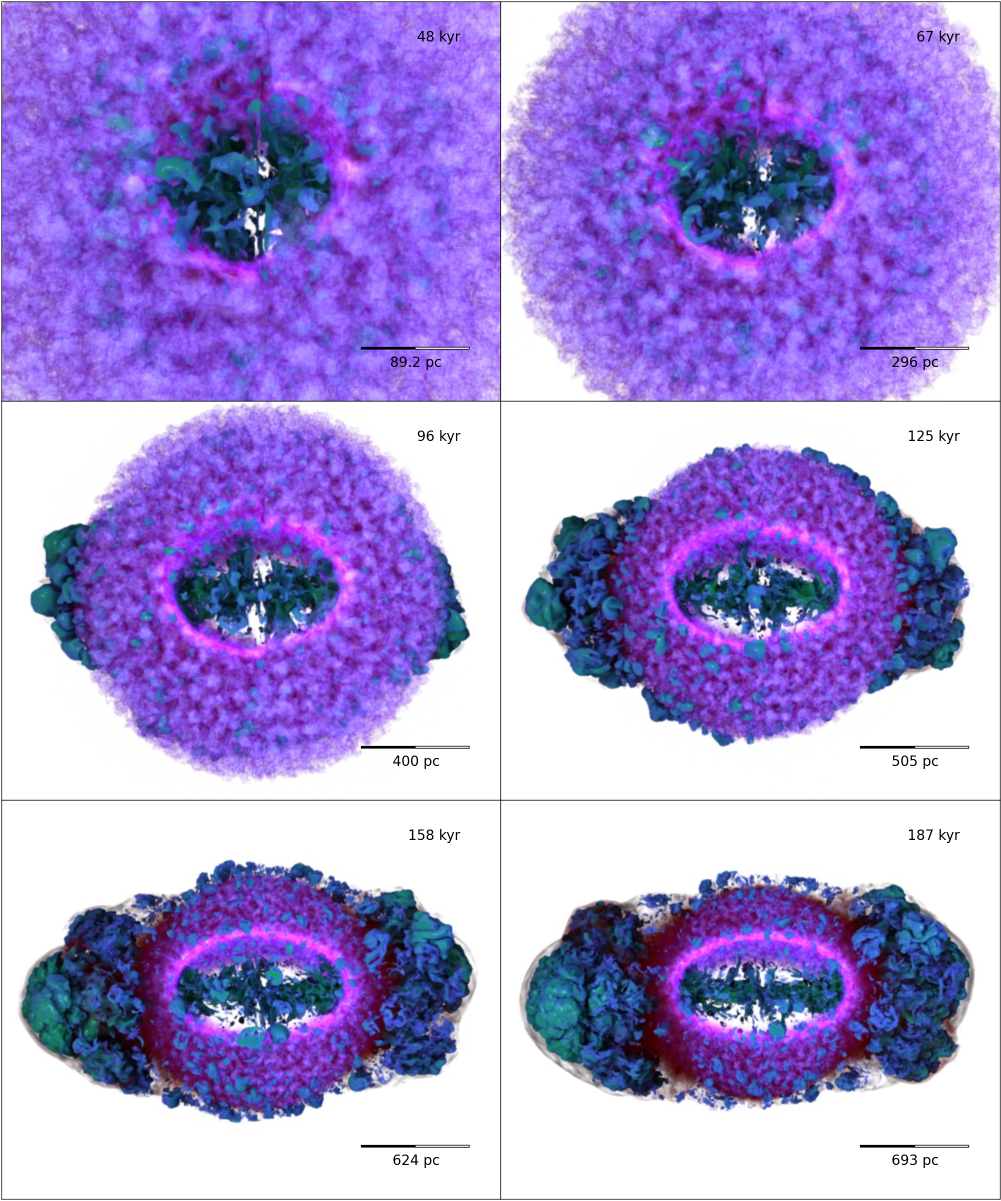}
  \caption{Volume render of the density of the jet plasma and clouds for run D$^\prime$. The jet plasma is textured in bluish green and the clouds in purple. The forward shock outlining the jet-blown energy bubble is seen in translucent grey. An oval excavation is made in the visualization of the clouds in order to show the jet plasma flow within. The view moves outward from the core of the galaxies as the bubble of jet plasma expands, and the physical (projected) size is indicated by scale bars on the bottom right in each panel. The simulation data is reflected about $x=0$ and the left side is rotated by $180^\circ$ about the jet axis to show a back view of the simulation. See the electronic edition of the Journal for a color version and an mpeg animation of this figure.}
  \label{f:vol}
\end{figure*}

We conducted 15 new simulations to study new regions in the space spanned by parameters that describe the distribution of warm phase material in our simulations as described in \S\ref{s:params}. Figure~\ref{f:dens} shows density maps of three selected new simulations with lower filling factor and differing maximum cloud sizes to those of previous simulations. To obtain a three dimensional impression of the interactions between the jet and the clouds we show a volume render of the density of both components from one of our simulations in Figure.~\ref{f:vol}. The jet plasma is textured in bluish green and the clouds in purple. The forward shock outlining the jet-blown energy bubble is seen in a translucent grey. An oval excavation is made in the visualization of the clouds in order to show the jet plasma flow within. 

The global evolution of the simulations was described by \citet{wagner2011a}. A key feature of the jet-ISM interactions is that whatever the initial narrowness of the jet, the jet flow is broadened by the interaction with the first cloud. The secondary jet streams flood through the porous channels of the two-phase ISM and a quasi-spherical jet-driven bubble sweeps over the entire bulge region. The feedback operates isotropically, without depending on the initial width or collimation of the jet, and the clouds at all position angles in the galactic halo are dispersed to high velocities.

Let $\MBH$, $\mprot$, $c$, $\Thomsonx$, and $\sigma_{100}$ be the black hole mass, the proton mass, the speed of light, the Thomson electron scattering cross section, and the velocity dispersion in units of $100\kms$, respectively. We also define the ratio of jet power to Eddington luminosity (the Eddington ratio) to be $\eta=\Pjet/\Ledd$, and $\phiw$, $\rho$, and $\vrad$ as the warm phase tracer (mass fraction in a cell), density, and radial velocity, respectively. A convenient measure of the efficiency of feedback is the density-averaged radial outflow velocity, $\vrw=\sum_{l=1}^N \phiw \, \rho \, \vrad / \sum_{l=1}^N \phiw \, \rho$, relative to the velocity dispersion of a galaxy's bulge as predicted by the \msigma{} relation \citep{silk1998a,king2005a}. Defining the black hole mass in terms of the Eddington ratio  $\MBH=4\pi{}G\mprot\Pjet{}c/\eta{}\Thomsonx$ and using the \msigma{} relation found by \citet{tremaine2002a} we express the velocity dispersion as 
\begin{equation}
\sigma_{100} = 1.0\eta^{-1/4} \, \Pjetff^{1/4}
\label{e:sigma}
\end{equation}
where $\Pjetff$ is the jet power in units of $45\ergs$. When $\vrw>\sigma$,  the jet-ISM interactions result in sufficient feedback of momentum and energy to establish a highly dispersed distribution of cold and warm gas within the core of the galaxy. The advantage of scaling the jet power by the Eddington luminosity is that it allows us to relate a given simulation to the conditions for feedback set by the \msigma{} relation in a galaxy with a SMBH mass according to the value $\eta$. We may scale the jet power arbitrarily because the simulations do not depend on the gravitational field for a particular $\MBH$ or bulge mass, $\Mbulge$.

\begin{figure}
  \figurenum{3}
  \includegraphics[width=\linewidth]{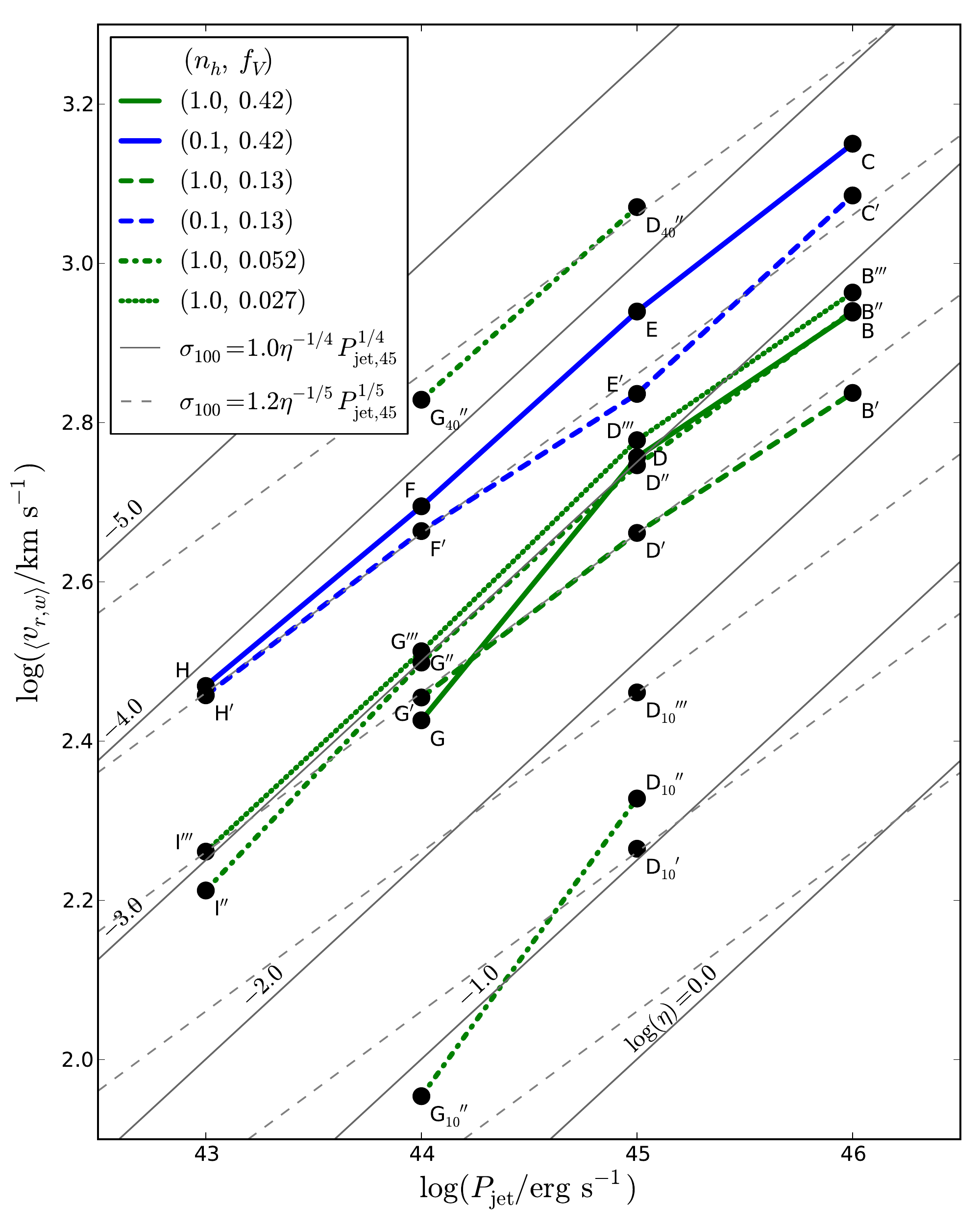}
  \caption{Maximum mean radial velocity of clouds, $\vrw$, against jet power for the simulations B -- I$^{\prime\prime\prime}$ of Table~\ref{t:runs}. The solid and dashed grey lines are loci of constant $\eta$, the ratio of jet power to Eddington luminosity, for \msigma{} relations with powers 4 and 5, respectively. The line colors indicate different hot-phase densities and the line styles represent different filling factors, as indicated in the legend. See the electronic edition of the Journal for a color version of this figure.}
  \label{f:v-P}
\end{figure}

The maximum values of $\vrw$ during the run as a function of jet power for all 28 simulations that include the warm phase are shown in Fig.~\ref{f:v-P}, which updates Fig.~5 in WB11. Points of constant hot phase density and filling factor are connected along increasing jet power with lines of specific colors and style according to the legend. The slanted grey lines in Fig.~\ref{f:v-P} represent the loci of the velocity dispersion along lines of constant $\eta$ (as determined by Eqn.~\ref{e:sigma}). The dashed grey lines in the figure represent a different \msigma{} relation of $\MBH\propto\sigma^5$, which is more in agreement with the recent results by \citet{graham2011a}, especially for core galaxies \citep{graham2012a}. The locus in this case is $\sigma_{100} = 1.2 \eta^{-1/5} \Pjetff^{1/5}$. Using either relation, one may then compare the points of the simulations with values for the velocity dispersion predicted by the \msigma{} relation for a given value of $\eta$. If a point lies above an isoline for $\eta$, then feedback by a jet of that power $\Pjet$ in a galaxy with Eddington limit $\Pjet/\eta$ is effective. Conversely if a point lies below an isoline for $\eta$, then feedback is not effective. Equivalently, the point itself marks a critical value of $\eta$, $\etacrit=(\Pjet/\Ledd)_\mathrm{crit}$, below which feedback ceases to be effective in galaxy with a SMBH of mass $\MBH=4\pi{}G\mprot\Pjet{}c/\eta{}\Thomsonx$.

As observed in previous simulations, the velocities attained by clouds match those observed of outflows in radio galaxies \citep{morganti2005a, holt2008a, nesvadba2006a, nesvadba2008a, nesvadba2010a, lehnert2011a, dasyra2011a, guillard2012a, torresi2012a}. The dense cores of the clouds in our simulations are accelerated to a few $100\kms$, while the diffuse ablated material is accelerated to several $1000\kms$. We discovered that the feedback efficiency of the relativistic jet on the warm phase ISM increases with increasing jet power, decreasing mean ISM density, and increasing filling factor, although only two values for the filling factor, $\fvol=0.42$, and $\fvol=0.13$, were studied.

Within the new range of parameter space, the main conclusions reached in WB11 remain valid; feedback is effective in systems in which the jet power is in the range $\Pjet=10^{43}$ -- $10^{46}\ergs$ and $\eta>\etacrit$. Furthermore, we find that, the maximum density weighted radial outflow velocity of clouds, $\vrw$, or equivalently, the critical Eddington ratio of the jets, $\etacrit$, depends weakly on filling factor, but strongly on the maximum size of clouds in the galaxy bulge. The overall lower limit $\etacrit\gtrsim10^{-4}$ for efficient feedback found by WB11 is only slightly reduced for galaxies containing small cloud complexes ($\Rcmax\lesssim10\pc$, $\kmin=40\invkpc$) but jets with Eddington ratios of $\etacrit=10^{-2}$ -- $10^{-1}$ are required if cloud complexes are large ($\Rcmax\gtrsim50\pc$, $\kmin=10\invkpc$).

\begin{figure}
  \figurenum{4}
  \begin{center}
  \includegraphics[width=\linewidth]{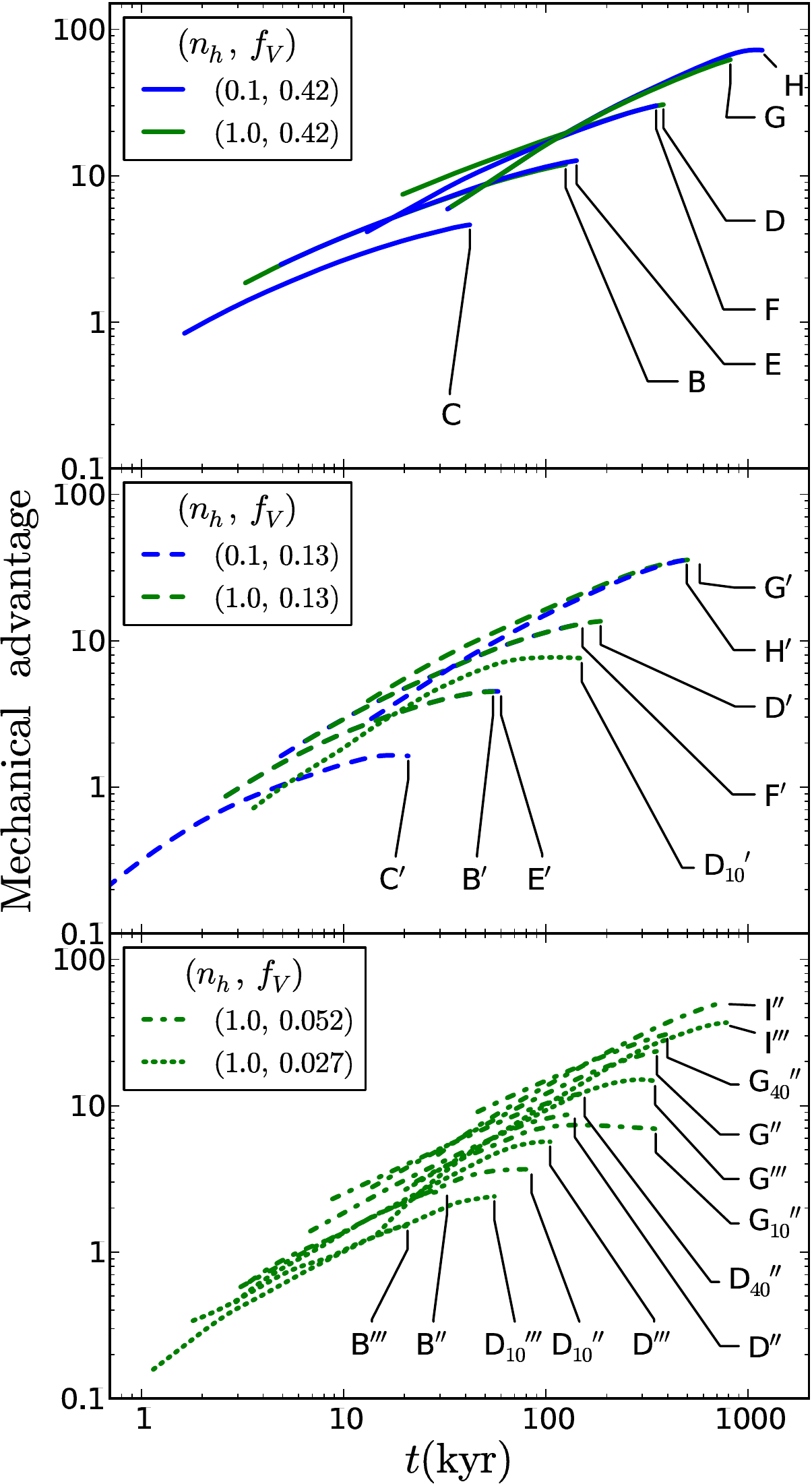}
  \end{center}
  \caption{Mechanical advantage versus time for all 28 runs including clouds. The top, middle, and bottom panels shows runs for which $\fvol=0.42$, $0.13$, and $0.052$ or $0.027$, respectively. The mechanical advantage here is defined as the total outward radial momentum of clouds at time $t$ divided by the total momentum delivered by the jet up to time $t$. The mechanical advantage in all simulations $\gg1$ indicating strong momentum coupling in the energy-driven regime. See the electronic edition of the Journal for a color version of this figure.}
  \label{f:madv}
\end{figure}

WB11 found that the jet-ISM interactions, despite the porosity of clouds and the radiative losses of shock-impacted clouds, exhibit a high mechanical advantage, meaning that substantial momentum transfer from the jet to the clouds occurred through the energy injected by the jet. We define the mechanical advantage in our simulations at a given time as the ratio of the total radial outward momentum carried by clouds to the total momentum delivered by the jet up to that time. Figure~\ref{f:madv} shows the curves for the mechanical advantage as a function of time for all 28 simulations including a warm phase. For all simulations the mechanical advantage is much greater than unity. Most curves fall closely on top of each other along a narrow band up to at least $1\Myr$.

\begin{figure}
  \figurenum{5}
  \begin{center}
    \includegraphics[width=\linewidth]{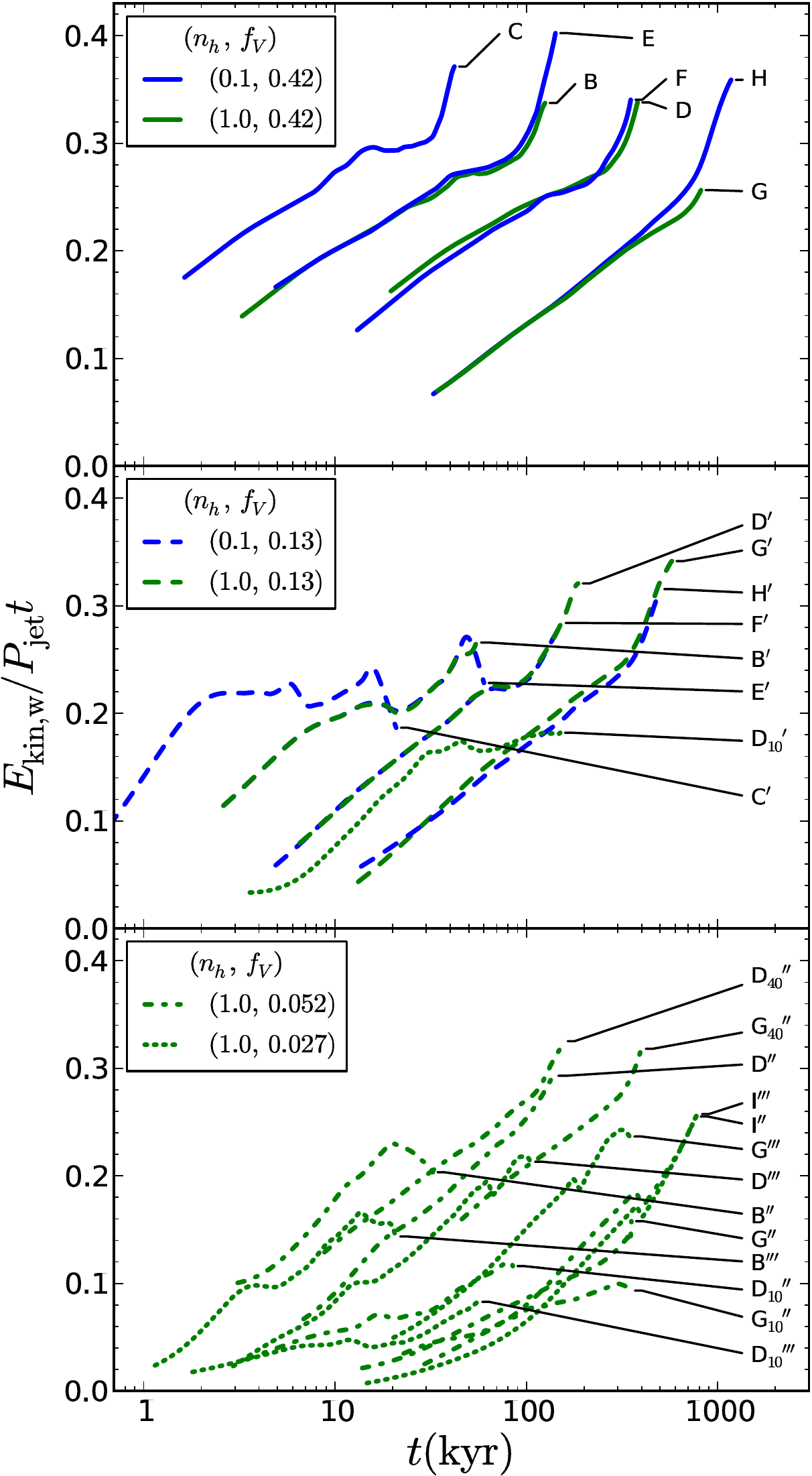}
  \end{center}
  \caption{Fraction of jet energy going into kinetic energy of the warm phase as a function of time for all 28 simulations containing a warm phase. The top, middle, and bottom panels shows runs for which $\fvol=0.42$, $0.13$, and $0.052$ or $0.027$, respectively. For all simulations, $0.4\gtrsim E_\mathrm{w,kin}/\Pjet\gtrsim0.1$ although the maxima and the time taken to reach the maxima depend on the jet power and ISM parameters. See the electronic edition of the Journal for a color version of this figure.}
  \label{f:eeff}
\end{figure}

The high mechanical advantage generally leads to a high fraction of jet energy transferred to kinetic energy of the warm phase. In Fig.~\ref{f:eeff} we show the evolution of the ratio of kinetic energy in clouds to injected jet energy, $E_\mathrm{kin,w}/\Pjet t$, as a function of $t$ for all 28 simulations including a warm phase. In all cases that fraction is high, reaching $\sim0.1$ -- $0.4$, with details depending on jet power and ISM properties. The details of the dependence on feedback efficiency on ISM properties are given in the next two sections and the physics of how the high mechanical advantage is sustained and the energy transfer occurs are investigated in \S\ref{s:ram}.

\subsection{Dependence on filling factor}\label{s:fvol}

\begin{figure}
  \figurenum{6}
  \includegraphics[width=\linewidth]{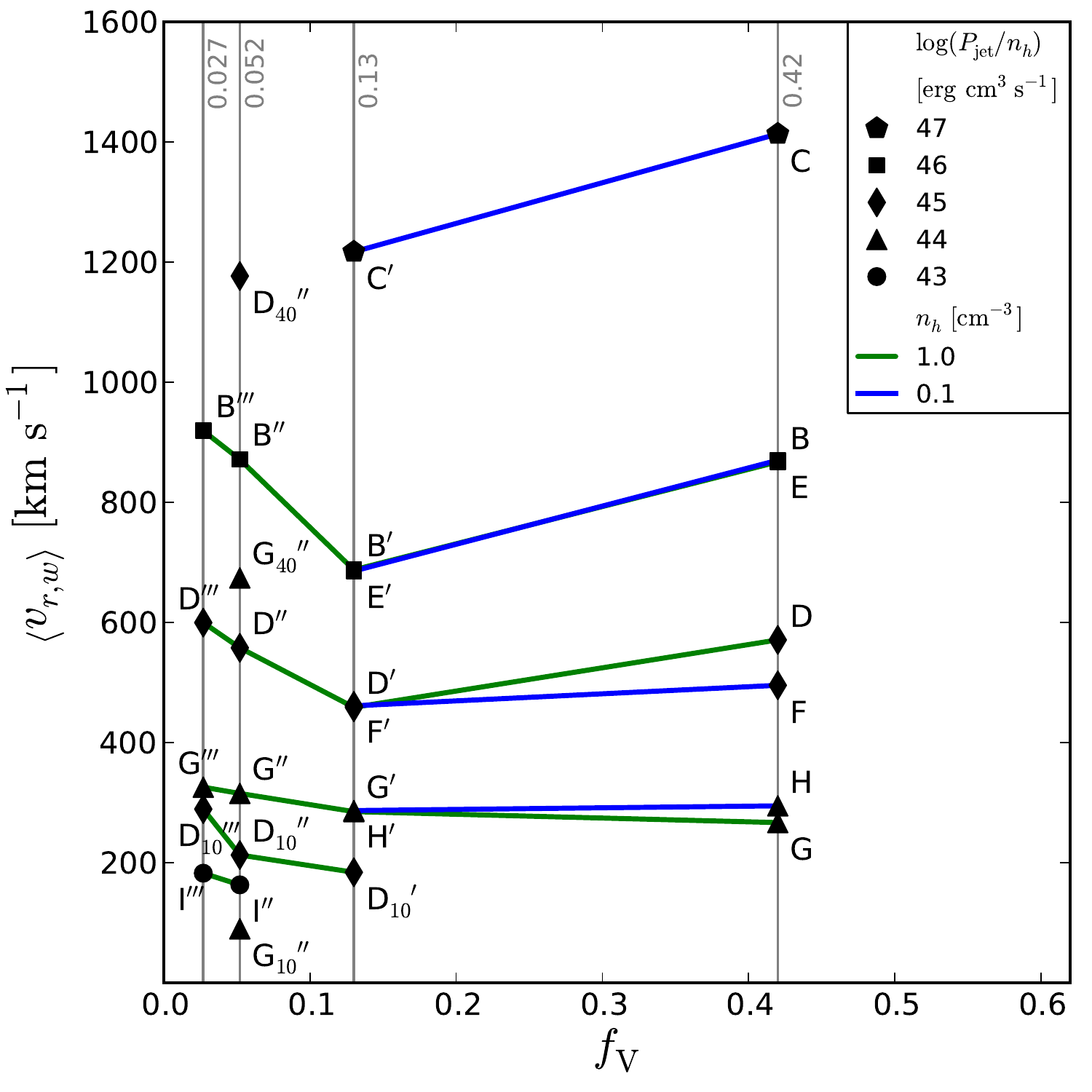}
  \caption{Maximum mean radial velocity of clouds, $\vrw$, versus cloud volume filling factor for the simulations B -- I$^{\prime\prime\prime}$ of Table~\ref{t:runs}. The line colors indicate different hot-phase densities and the marker styles group simulations with equal values of $\Pjet/\nhot$, as indicated in the legend. See the electronic edition of the Journal for a color version of this figure.}
  \label{f:v-fv}
\end{figure}

Figure~\ref{f:v-fv} shows the maximum values of $\vrw$ reached in the simulations as a function of $\fvol$. The markers denote simulations with equal values of $\Pjet/\nhot$, as indicated in the legend. The lines of a given color connect simulations of equal power, also indicated by the label letter, and the line color indicates the hot phase density. Apart from the cases of different $\kmin$ in the D-series of runs, simulations grouped by connected lines, therefore, also indicate runs with equal values of $\Pjet/\nhot$.

In general, the dependence of $\vrw$ on filling factor is weak and non-monotonic. In the B-series, D-series, D$_{10}$-series, G-series, and I-series of the simulations (see table~\ref{t:runs} for nomenclature), we observe that for $\fvol\gtrsim0.1$, lower filling factors decrease the feedback efficiency, while for $\fvol\lesssim0.1$, lower filling factors increase the feedback efficiency. The reason for the weak dependence and the non-monotonicity is the competing effects of the cloud ablation rate and jet plasma confinement time. On the one hand, smaller filling factors increase the volume available for the jet plasma to flood through, and thereby reduce the confinement time, which reduces the impulse delivered to the clouds over the confinement time. On the other hand, smaller filling factors increase the mass of ablated material relative to the total mass of a cloud, because the ablation rate is proportional to the cloud surface area and the mass is proportional to the cloud volume, which decreases faster than the former for decreasing filling factor. When lowering the filling factor in the range $\fvol\gtrsim 0.1$ the effect of reduced plasma confinement time dominates over the effect of increased fractional cloud ablation and results in lower mass-averaged outflow velocities. In the range $\fvol\lesssim 0.1$, the increased cloud ablation rate dominates over the reduced plasma confinement time when reducing $\fvol$, leading to higher mass-averaged outflow velocities. Over the range of values for the filling factor studied here, these two effects counteract one another, and the dependence of $\vrw$ on $\fvol$ for constant $\Pjet/\nhot$ remains weak. 

The mechanical advantage (Fig.\ref{f:madv}) is slightly reduced for systems with lower filling factor down to $\fvol=0.027$, but the dependence of the efficiency of transfer of jet energy to kinetic energy of the warm phase (Fig.\ref{f:eeff}) on warm-phase filling factor parallels the weak (non-monotonic) dependence of the maximum outflow velocity on filling factor.

Note that, by reducing the filling factor, we are also reducing the total mass of the warm phase. in contrast to this, we may keep the total mass and filling factor the same but change the maximum size of clouds by varying $\kmin$. The results for this are shown next.

\subsection{Dependence on maximum cloud sizes}\label{s:kmin}

Let us look at the D-series of runs, for which we have varied the maximum size of clouds by varying $\kmin$. The values of $\kmin$ are denoted by the subscript of the run labels in Fig.~\ref{f:v-P}.

\begin{figure}
  \figurenum{7}
  \includegraphics[width=\linewidth]{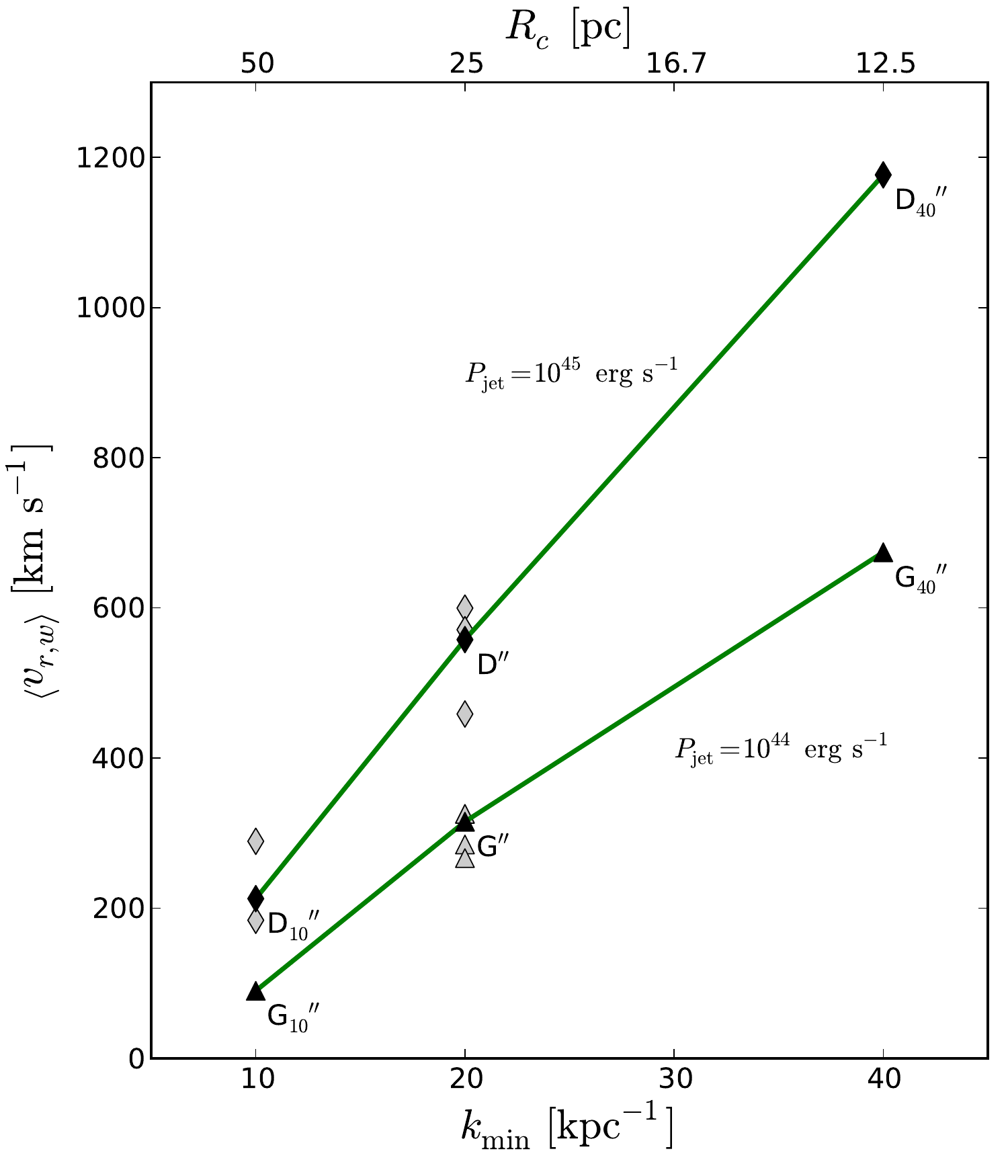}
  \caption{Maximum mean radial velocity of clouds, $\vrw$, versus minimum sampling wave number, $\kmin$ (and corresponding maximum cloud size, $\Rcmax$), for runs in the D-series (diamond points, $\Pjet=10^{45}\ergs$) and G-series (triangular points, $\Pjet=10^{44}\ergs$). The runs with filling factor $\fvol=0.052$ are marked with black markers, labeled, and connected with lines and show the variation of $\vrw$ with $\kmin$. The grey markers clustering around a black marker are runs differing only in filling factor. See the electronic edition of the Journal for a color version of this figure.}
  \label{f:v-km}
\end{figure}

In Fig.~\ref{f:v-km}, we plot the sequences D$_{10}^{\prime\prime}$, D$^{\prime\prime}$, D$_{40}^{\prime\prime}$, and G$_{10}^{\prime\prime}$, G$^{\prime\prime}$, G$_{40}^{\prime\prime}$ against $\kmin$ with black markers, labels, and connected by a line. The grey, unlabelled markers are other runs with varying filling factors but otherwise identical parameters. The sequences in this figure, but also those in both Figs.~\ref{f:v-P} and \ref{f:v-fv}, show that the dependence of mean velocity on the maximum cloud size in a simulation is very strong, and that is much stronger than the dependence on filling factor. 

By halving the size-scale of clouds from $\kmin=20\invkpc$ (D$^{\prime\prime}$) to $\kmin=40\invkpc$ (D$_{40}^{\prime\prime}$), the feedback provided by the jet accelerates the clouds to a velocity a factor of two greater, from $600\kms$ to $1200\kms$. Doubling the cloud sizes from $\kmin=20\invkpc$ to $\kmin=10\invkpc$ decreases the maximum cloud velocities reached in the simulation by a factor of 3, from $200\kms$ to $600\kms$. $\etacrit$ is therefore more sensitive to the maximum sizes of clouds than the volume filling factor of clouds. Moreover, the scaling between $\kmin$ and $\vrw$ is nearly linear between $\kmin=20\invkpc$ and $\kmin=40\invkpc$, and somewhat steeper than linear between $\kmin=10\invkpc$ and $\kmin=20\invkpc$.

The reason for the strong cloud-size dependence and linear scaling is that changing the cloud sizes at constant filling factor changes the rate of ablation relative to the total cloud mass without changing the jet plasma confinement time. This is because only the amount of surface area exposed to ablation relative to the volume of a cloud changes. Since $\kmin\propto\Rcloud^{-1}$, where $\Rcloud$ is the cloud radius, the ratio of surface area to volume of a cloud scales linearly with $\kmin$. For higher $\kmin$, the rate of ablation relative to the total mass of clouds increases, while the confinement time of the jet plasma does not change compared to runs with different $\kmin$ but identical $\fvol$. This allows for a far higher fraction of warm phase mass to be accelerated to higher velocities, increasing the maximum value of $\vrw$ reached in the run.

An equivalent statement to the above explanation uses the concept of a jet-cloud \lq\lq{}interaction depth\rq\rq{}, $\taujc$, for a given distribution of clouds with varying $\kmin$. In analogy to optical depth, the effective interaction depth may be written $\taujc = n_c \pi \Rcmax^2 \Rbulge$, where $n_c$ is the number of clouds per unit volume (the number density of clouds), and $\Rbulge$ is the radius of the region in the bulge which contains clouds. The clouds may be thought of as $N$ scattering centers with a cross-section $\pi\Rcmax^2$ randomly distributed in the volume $(4\pi/3)\Rbulge^3$, and the interaction depth may be thought of as a measure of the average number of jet-cloud interactions any jet stream starting from the origin, including trajectories along secondary streams, will experience. This formulation of an interaction depth is indeed relevant because, as we demonstrate in \S\ref{s:ram}, the jet streams carrying entrained hot and warm phase material are directly responsible for the acceleration of clouds through their ram pressure. The total number of clouds in the bulge is $N=\fvol\Rbulge^3/\Rcmax^3=n_c\Rbulge^3$. Therefore, the number density of clouds is $n_c=\fvol/\Rcmax^3$, and the interaction depth is $\taujc=\pi\fvol(\Rbulge/\Rcmax)=\pi\fvol\kmin$. Hence, for a fixed $\fvol$, $\taujc\propto\kmin$.

The linear relation between the ratio of surface area to volume of a cloud and $\kmin$, or equivalently, the linear relation between $\taujc$ and $\kmin$, leads to a linear relation between $\kmin$ and $\vrw$, which is seen between $\kmin=20\invkpc$ and $\kmin=40\invkpc$. It is, however, not clear whether one may extrapolate this relation to larger cloud complexes with sizes characteristic of giant molecular clouds (GMC), say of order several $100\pc$, given that the scaling between $\kmin=10\invkpc$ and $\kmin=20\invkpc$ is steeper than linear. A possible reason for the steepening at larger $\Rcmax$ is that the larger inter-cloud voids cause a decollimation of the jet streams leading to less efficient momentum transfer. The scaling may also be affected by resolution limitations to capturing the fractal surface of clouds, and by statistical variations in the decreased number of jet-cloud interactions for small $\kmin$. It is difficult to predict the feedback efficiency with respect to GMC with scales of order $100\pc$ from our simulations because these are generally not spherical and the effective interaction cross-section depends on orientation with respect to the jet streams. The simulations by \citet{sutherland2007a} and \citet{gaibler2011a} show that, if the molecular gas is distributed as a large coherent complex in a disc-like geometry, the coupling between jet and ISM in terms of negative feedback through gas expulsion is weak. Observations of some gas-rich radio galaxies indicate that the molecular gas is not coupled as strongly into outflows with the jet as the neutral or ionized material \citep[][]{ogle2010a,guillard2012a}, although 4C 12.50 is a prominent exception \citep{dasyra2011a,dasyra2012a}.

The explanations given here also apply to the influence of cloud sizes on the mechanical advantage and energy transfer efficiency from jet to warm phase in these systems. Both the mechanical advantage (Fig.\ref{f:madv}) and the energy transfer efficiency (Fig.\ref{f:eeff}) are significantly greater in systems with smaller cloud sizes.

The sense in which the fractional cloud dispersal rate depends on cloud sizes is the same as the dependence of the conditions for star formation in a cloud on its size, in that, the larger a cloud the more likely it is to collapse due to an external pressure trigger. Thus, whether jet mediated feedback induces or inhibits star-formation is a sensitive function of the statistics of the warm phase distribution, in particular its size distribution.

\subsection{The expansion rate of the quasi-spherical bubble}\label{s:bubble}
In this section, we determine the departure of the outflow energetics from that of an energy-driven bubble as functions of warm phase parameters. Because our simulations include radiative cooling and a porous two-phase ISM, we expect the energetics of the bubble that sweeps up the ISM imparting momentum and energy to the clouds to lie between the energy-driven and momentum-driven limits. We discuss momentum-driven and energy-driven outflows in relation to work in the literature separately in \S\ref{s:more}.
\begin{figure*}
  \figurenum{8}
  \centering
  \includegraphics[width=0.8\textwidth]{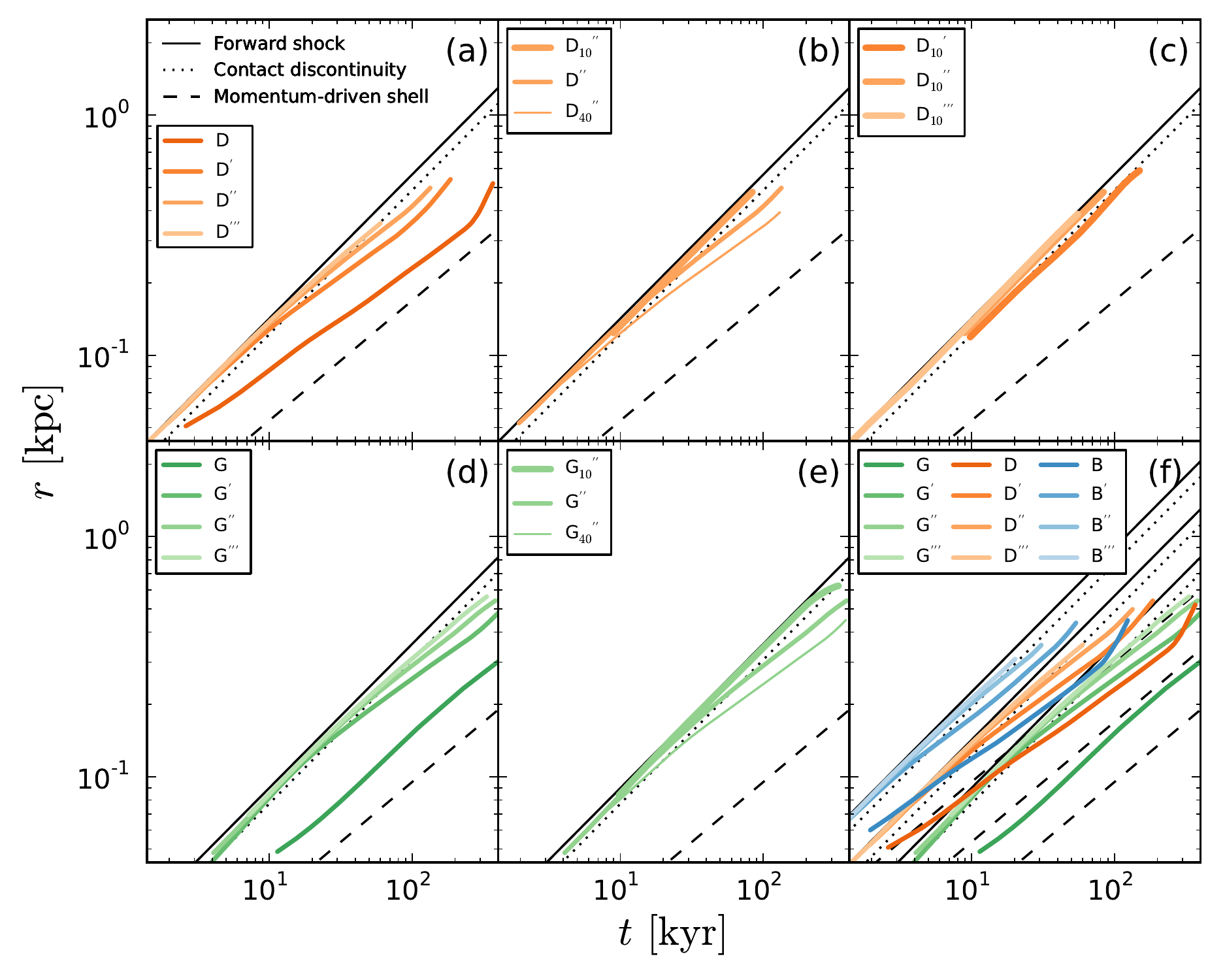}
  \caption{The spherically equivalent bubble radius as a function of time for simulations with different ISM parameters in the G-series ($10^{44}\ergs$), D-series ($10^{45}\ergs$), and B-series ($10^{46}\ergs$) of runs. The runs have identical ISM hot phase density ($\nhot=1.0\cmq$), but differ in jet power, warm phase filling factor, and maximum cloud sizes. The solid black line and the dotted black line are position of the forward shock and the position of the contact discontinuity in an energy-driven bubble \citep{weaver1977a, bicknell1996a}. The dashed black line is the location of the thin shell in a momentum-driven bubble \citep[e.g.]{dyson1984a}. \textit{a}: Bubbles driven by jet feedback in lower filling factor environments evolve closer to classical energy-driven bubbles. \textit{b}: Bubbles driven by jet feedback in halos with larger cloud complexes (at constant filling factor) evolve closer to classical energy-driven bubbles. \textit{c}: Same as \textit{a} except for $\kmin=10\invkpc$. \textit{d}: Same as \textit{a}, but for the G-series of runs. \textit{e}: Same as \textit{b}, but for the G-series of runs. \textit{f}: The dependence of the bubble expansion rate on ISM parameters is similar for all jet powers. See the electronic edition of the Journal for a color version of this figure.}
  \label{f:bubble}
\end{figure*}

Figure~\ref{f:bubble} contains six panels showing the evolution of the bubble radius with time for different runs in the G-series, D-series, and B-series. We defined the bubble radius to be the radius of a hemisphere whose volume is equivalent to that swept up by the pressure bubble in the simulation. In each panel, the solid black line and the dotted black line represent the theoretical, self-similar, spherically symmetric evolution of the forward shock radius and contact discontinuity, respectively, of an energy-driven bubble (wind) in a uniform medium in Stage 1, as defined by \citet{weaver1977a} \citep[see also \S6][]{bicknell1996a}. That stage represents an adiabatically expanding bubble with constant injection power, which in our case is $\Pjet$. The solutions of the first stage are applicable here because radiative losses, although they improve the structural integrity of clouds and their survival time \citep{cooper2009a}, are energetically unimportant in our simulations. The forward shock and discontinuity evolve according to $R_2=0.88(\Pjet/\nhot)^{1/5}t^{3/5}$ and $R_C=0.86 R_2$, respectively. The location of the thin shell in the momentum-driven limit of a bubble expanding in a uniform medium of mass density $\rhohot$ is delineated by the dashed black line, and given by the equation $R_\mathrm{shell}=\sqrt{3/2}(\dot{p}/\rhohot)^{1/4}t^{1/2}$, where $\dot{p}$ is the momentum injection rate \citep{dyson1984a}.

Without focusing on a particular panel in Fig.~\ref{f:bubble}, we note that the bubble radius in some runs follows that of the theoretical prediction for an energy-driven bubble closely, while in others the bubble radius initially increases more slowly than the rate predicted by theory. A slight deceleration can even be seen in some runs as the bubble is increasingly mass-loaded by warm phase material. In a few runs a return toward the theoretical line is visible, with gradients steeper than the theoretical limit for an energy-driven wind of given injection power. This happens because the medium inside the pressure bubble while the jet plasma is confined by clouds is at a higher pressure than that of a bubble that is expanding in a homogeneous atmosphere with ambient density $\nhot$. During the breakout phase of the jet from the region filled with clouds, the jet plasma bursts out of the outermost porous channels and momentarily fills volumes at a faster rate than a bubble that was not impeded and confined for some duration by a porous, dense distribution of clouds.

The first panel ({\it a}) shows four runs of differing filling factor from the D-series, for which $\kmin=20\invkpc$. A bubble evolving in a system with larger filling factor expands more slowly. As we decrease the filling factor, the deviation from an energy-driven bubble become smaller. We see the same behaviour for the runs in the G-series (panel {\it d}). The larger volume of channels available for the jet plasma to flood through, and the resulting smaller confinement time, is the dominant factor that defines the bubble expansion rate. The same trend is visible in the second panel for simulations of differing filling factor, for which $\kmin=10\invkpc$ (panel {\it c}), although the effect is much weaker. This is the result of the confinement times and mass loading from hydrodynamic ablation ceasing to vary much as the maximum cloud sizes increases.

The expansion rate profiles for runs with differing maximum cloud sizes, but equal filling factor at $\fvol=0.052$ is shown in panels ({\it b}) and ({\it e}) for the D-series and G-series, respectively. The expansion rate of the bubble deviates increasingly from the theoretical rate with decreasing maximum cloud sizes. The reason for this is the increased mass ablation rate relative to the total mass of clouds, as described in \S\ref{s:kmin}. Smaller maximum cloud sizes for the same porosity lead to higher mass-loading rates and decreased expansion rates of the bubble.

While feedback efficiencies are mainly sensitive to the maximum cloud sizes, the deviation of the bubble expansion rate from a theoretical energy-driven rate is sensitive to both the maximum cloud sizes and filling factor. Within the parameter range studied here the deviation depends more strongly on filling factor than maximum cloud size. For systems containing large clouds with small filling factors, the bubble evolution approaches that of an ideal energy-driven bubble. Thus, it would seem that in this limit, these results encourage a sub-grid AGN feedback prescription in cosmological models, in which energy is injected isotropically into a small region, even if the multiphase ISM conditions in the cores of gravitational potentials are not adequately resolved. However, this limit is not the same as that which leads to the most efficient cases of negative AGN feedback. The latter is attained in the limit of small filling factors and small cloud sizes. A distribution of larger clouds, instead, may lead to positive feedback, e.g., pressure-triggered star-formation.

Cosmological SPH models commonly invoke negative AGN feedback in a single-phase ISM, which essentially corresponds to the hot phase in our simulations. The heating rate of the hot phase will therefore likely always be accurately captured in these models if the filling factor is smaller than $\sim 0.1$. The dispersal of warm and cold gas, on the other hand, requires $\kmin\gtrsim20\invkpc\;(\Rcmax\lesssim25\pc)$. Since the bubble expansion rate does not depend very strongly on the maximum size of clouds, however, we assert that negative feedback as implemented in cosmological SPH models in a single-phase ISM is consistent with negative feedback in our simulations with a two-phase ISM if the warm phase filling factor is less than $\sim 0.1$ and the largest cloud complexes are smaller than $\sim25\pc$. In this regime, the embedded warm-phase material is accelerated nearly isotropically to the bubble expansion speed within the dynamical time of the bubble, ensuring that the negative feedback affects both phases, while the bubble remains approximately energy-driven. This conclusion is independent of jet power.

\subsection{Energy- or momentum-driven?}\label{s:more}

The theories of a momentum-driven wind developed by \citet{fabian1999a}, \citet{king2003a}, and \citet{murray2005a} naturally predict $\MBH\propto\sigma^4$. The theory of an energy-driven wind put forward by \citet{silk1998a} predicts the relation $\MBH\propto\sigma^5$. These two relations and their normalizations are limiting cases for outflow velocities that can be reached in an outflow powered by $\Ledd(\MBH)$. In the former case, the outflow loses its internal energy through radiative processes (e.g. Inverse Compton cooling) and its dynamics is governed solely by momentum conservation, while in the latter case energy is fully conserved. The difference of 1 in the exponent of the relations is not surprising from dimensional arguments, since energy conservation entails a dependence on velocity squared as opposed to a linear dependence on velocity associated with momentum-driven flows. 

In Fig.\ref{f:v-P}, the solid lines of constant $\eta$ represent the limiting slope of a momentum-driven outflow and the dashed lines represent the limiting slope for an energy-driven outflow. The loci of the maximum values of $\vrw$ in our simulations with identical filling factor and ISM densities and $\kmin=20\invkpc$ cluster between $\log(\eta)=-2$ and $\log(\eta)=-4$ along narrow strips roughly parallel to $\eta$ isolines. The average gradient of the lines connecting the loci of $\vrw$ appears to lie between 1/4 and 1/5, indicating that the outflow is somewhere between momentum and energy-driven. 

In a two-phase medium, the determination of whether a bubble evolves in the momentum-driven regime or energy-driven regime depends on how one compares the evolution with that for the case of a smooth, single-phase ambient medium. This, in turn, depends on what feedback criteria one is interested in. In \S\ref{s:bubble} we saw that a bubble evolves in the energy-driven regime as long as the warm phase volume filling factors were not larger than 0.1. In this regime, the radial heating rate of the hot phase (but not necessarily the warm phase) is well described by that of an energy-driven bubble. The suppression of star-formation in existing clouds is effective only if the additional constraint of $\Rcmax\gtrsim25\pc$ is satisfied, because the clouds are then efficiently ablated, heated, and dispersed. Only in this regime can the entire two-phase medium be considered as expanding approximately in the energy-driven limit, because the warm phase is accelerated to the bubble expansion speed within the dynamical time of the bubble.

One expects energy-driven and momentum-driven outflows to differ in their kinetic power required to achieve feedback. For example, consider the predictions of the \citet{silk1998a} model, which employs the same condition for feedback as we have used in our work to derive the \msigma{} relation, namely that the outflow velocity exceed the host galaxy's bulge velocity dispersion. The normalization to the derived \msigma{} relation contains a wind efficiency parameter $\fw=\mdotout\vout^2/\Ledd$, where $\vout$ is the outflow velocity, and $\mdotout$ is the mass injection rate of the wind. The values of $\fw$ can be compared to those of $\eta$ in this work, because $\Ekinw/\Pjet\sim0.1$ -- 0.4 in our simulations (see Fig.~\ref{f:eeff}). From observational estimates of the energetics of AGN outflows, one would expect this factor to be of order 0.001 -- 0.01 \citep[e.g.][]{mckernan2007a,moe2009a,tombesi2010b}, which is also found in disc-wind simulations \citep{kurosawa2009a,takeuchi2010a,ohsuga2011a}. This is the level at which cosmological SPH simulations typically inject energy to model AGN feedback \cite[e.g.][]{di-matteo2005a,okamoto2008a,booth2009a}. However, comparing the normalization of the observed \msigma{} relation to that derived by Silk \&{} Rees, one obtains $\fw\sim7\times10^{-6}\fgas$, where $\fgas$ is the gas fraction in the dark matter halo, indicating that in the spherically symmetric energy-driven regime, low Eddington ratio outflows are sufficient to significantly disperse gas. This is also the result found in some semi-analytic models \citep[e.g.][]{croton2006a} where heating by AGN with small Eddington ratios suffices to suppress star-formation in massive galaxies and offset cooling in clusters. A possible reconciliation for these inconsistencies is that the observed outflows are momentum-driven. For example, a momentum-driven wind maintained by the photon momentum flux of an Eddington limited accretion flow ($\Ledd/c$) requires $\fw=\vout/c\sim\few\times10^{-3}$ to accelerate a spherically symmetric shell of swept up ISM to a velocity $v=\sigma$. 

As mentioned however, one must be careful about the types of feedback criteria one is comparing in the various studies (see also \S\ref{s:othf}), and how simulations partition the injected energy into thermal or kinetic energy in subgrid feedback prescriptions. The theories by \citet{silk1998a}, \citet{fabian1999a}, \citet{king2003a}, and \citet{murray2005a} aim to explain the \msigma{} relation. To inhibit star-formation within a galaxy it suffices to merely heat or ionize the dense gas. To offset cold gas accretion and avoid late-time star-formation, more powerful outflows are required to heat the IGM. \citet{silk2010a} argue that, Eddington-limited AGN do not provide enough energy during their lifetime to generate momentum-driven or energy-driven winds that can unbind the gas from the galaxy potential. AGN jets are sometimes favoured in these cases. In cosmological SPH simulations, AGN feedback is usually implemented as thermal energy injection into particles near the core of galaxies, which effectively results in an energy-driven bubble with maximal mechanical advantage that heats the ISM and efficiently inhibits local star formation and cluster-scale cooling flows. The feedback requires relatively high injection rates of thermal energy, compared to a wind that is dominated by kinetic energy and can drive the bubble through ram pressure as well thermal pressure \citep{ostriker2010a}.

In this work, we found that the minimum value of $\eta$ required by a jet to disperse the warm phase to the velocity dispersion implied by the \msigma{} relation is $\eta\gtrsim10^{-4}$ and depends on the ISM density, filling factor, and cloud sizes. The regime of $\fvol\gtrsim0.1$ and $\Rcmax\lesssim25\pc$ comes closest to an energy-driven bubble of the entire two-phase medium, and, $\etacrit$ is accordingly small ($<10^{-4}$). This is not surprising since the limit $\kmin\rightarrow\infty$ is essentially a single-phase medium, and the surface area for ram pressure and thermal pressure to work on (and consequently the mechanical advantage) is maximised. Even though the bubble expansion rate depends quite sensitively on filling factor, $\etacrit$ does not. For $\Rcmax\gtrsim25\pc$ the clouds are not strongly ablated and accelerated by the bubble. The bubble containing mainly hot phase gas expands in a nearly energy-driven manner, but $\etacrit$ is large. These results for $\etacrit$, therefore, demonstrate that both filling factor and the size distribution of clouds need to be considered when assessing the efficiency (and type) of feedback, because these factors influence the degree to which the warm phase is incorporated in the outflow. Conversely, whether feedback is efficient or not cannot be uniquely determined by assessing whether an outflow is energy- or momentum-driven. 

In our simulations the bubble evolves near the energy-driven limit unless $\fvol\gtrsim0.1$. Other indications supporting this approximation are: 1) the mechanical advantage (Fig.~\ref{f:madv}) and the efficiency of energy transfer from jet to the warm phase (Fig.~\ref{f:eeff}) are high at all times; 2) For constant $\kmin$, the feedback efficiency scales roughly with $\Pjet/\nhot$ (Fig.~\ref{f:v-fv}), a characteristic parameter of energy-driven outflows. In the following section we investigate in detail how the warm-phase material is accelerated nearly isotropically to the bubble expansion speed within the dynamical time of the bubble, under the assumption that the bubble evolves in the energy-driven regime.

\subsection{Cloud acceleration through ram-pressure driving}\label{s:ram}

We have investigated in more detail the physical mechanisms that accelerate the clouds. As contained in the fluid equations (Eqns.~\ref{e:rel}), gradients in both ram pressure and thermal pressure contribute to momentum transfer from the jet plasma to the warm phase. The mechanical advantages measured in WB11 and here very high ($\gg1$) for all simulations (see Fig.~\ref{f:madv}).

We first show that the simple picture that the expansion velocity of the jet-driven bubble exerting a ram pressure on the clouds is responsible for the cloud velocity does not work. Careful inspection of the flow in the simulations, instead, shows a combination of other effects combining to provide a high mechanical advantage.

The analytic expression for a jet-blown bubble in the energy-conserving limit is not a bad approximation in this context as has been shown by our simulations. The radius $R_b$ of a bubble blown by a jet with power $\Pjet$ into a medium with ambient density $\rho_a$ is given as a function of time $t$ by
\begin{equation}
R_b = A t^{3/5}\;, \label{e:rbub}
\end{equation}
where
\begin{equation}
A = \left( \frac {125 \Pjet}{384 \pi \rho_a} \right)^{1/5}\;.
\end{equation}

The driving of a spherical cloud of mass $m_c = 4 \pi /3 R_c^3$ and radius $R_c$, to a velocity $v_c=\vexp$  via the ram pressure of the expanding bubble, which has rest mass density $\rho_b$ and expansion velocity $\vexp = (3/5) At^{-2/5}$ is described by the equation of motion:
\begin{equation}
m_c \td{v_c}{t} = C_D \rho_b \vexp^2 \times \pi R_c^2\;,
\end{equation}
where $C_D$ is the drag coefficient and $\pi R_c^2$ is the cross-sectional area of the cloud. The acceleration of a cloud is, thus,
\begin{equation}
\td{v_c}{t} = \frac{3 C_D}{4} \frac{\rho_b}{\rho_c} \frac{v_{\rm exp}^2}{R_c} \;.
\label{cloud_accn} 
\end{equation}

If we assume in this initial model that the density of the bubble is determined the jet mass flux $\Mdotjet$, then
\begin{equation}
\td{}{t} \left( \frac{4\pi}{3}  \rho_b R_b^3\right) = \Mdotjet\;,
\end{equation}
and, integrating, the bubble density is
\begin{equation}
\rho_b = \frac{3}{4 \pi} A^{-3} \Mdotjet t^{-4/5} \label{e:rhobjet}
\end{equation}
We can use equation~(\ref{cloud_accn}) to calculate an acceleration timescale for the cloud from
\begin{equation}
\frac{\vexp}{\tacc} = \frac{3}{4} C_D \left( \frac{\rho_b}{\rho_c} \right) \frac{\vexp^2}{R_c} \label{e:tacc1}
\end{equation}
and this implies that the acceleration timescale for clouds
\begin{equation}
\tacc = \frac{16 \pi}{9} C_D^{-1} \left( \frac{\rho_c}{\Mdotjet} \right) \frac{R_c}{\vexp} A^3 t^{4/5} \;.\label{e:tacc2}
\end{equation}
The acceleration timescale for clouds compared to the evolution time for the bubble is:
\begin{equation}
\frac {t_{\rm acc}}{t} = \frac{80\pi}{27} C_D^{-1} \left( \frac{\rho_c}{\Mdotjet} \right) R_c A^2 t^{1/5}  \;.
\label{t_acc}
\end{equation}

In order to evaluate $t_{\rm acc}$ we need to determine the jet mass flux. This can be determined from the jet power as follows. For a jet with relativistic enthalpy, w, velocity $c \beta$ and cross-sectional area $A_{\rm jet}$ the jet power is given by
\begin{equation}
\Pjet = \Gamma^2 w c \beta A_{\rm jet} - \dot M c^2  \;,
\label{e:energy_flux}
\end{equation}
where the mass flux, $\dot M$, for a jet with proper rest-mass density is given by
\begin{equation}
\dot M = \Gamma \rho c \beta A_{\rm jet} \;.
\end{equation}
The energy flux equation has two terms. The first term is the conventional relativistic form \citep{landau1987a} which incorporates the energy flux due to the rest mass energy plus the normal kinetic and internal energy terms. These contributions all originate from the relativistic enthalpy $w$. The second term subtracts off the rest mass energy flux since, in this context this is not useful energy. (Nuclear reactions are not involved.)

The ratio of the dynamic variables appearing in the energy and rest-energy fluxes is
\begin{equation}
\frac {\Gamma^2 v c \beta}{\Gamma \rho c^3 \beta} = \Gamma \frac {w}{\rho c^2} \;.
\end{equation}
Hence the energy flux is given by
\begin{equation}
\Pjet = \left( \Gamma \frac {w}{\rho c^2} -1 \right) \Mdotjet c^2 \;.
\end{equation}
We put
\begin{equation}
w = \rho c^2 + (\epsilon +p) \;,
\end{equation}
where $\epsilon$ is the internal energy density and $p$ is the pressure. This gives
\begin{equation}
\Pjet = (\Gamma-1) \Mdotjet c^2 \left[ 1 + \frac {\Gamma}{\Gamma-1} \frac {1}{\chi} \right] \;,
\end{equation}
where $\chi = \rho c^2 / (\epsilon+p)$. Therefore, the mass flux may be expressed in terms of the energy flux by:
\begin{equation}
\Mdotjet = \frac {1}{\Gamma-1} \frac {\Pjet}{c^2} \left[1 + \frac {\Gamma}{\Gamma-1} \frac {1}{\chi}  \right]^{-1} \label{e:mdotjet}
\end{equation}

For typical values, e.g., $\Gamma=10$, $\Pjet=10^{45}\ergs$, $\chi=1.6$, $C_D=1$, $\rho_c=1000$, $\rho_a=1$, and $R_c=25\pc$, it is obvious that this model does not work because $\tacc/t \gg 1$.

\begin{figure*}
  \figurenum{9}
  \begin{center}
    \includegraphics[width=\textwidth]{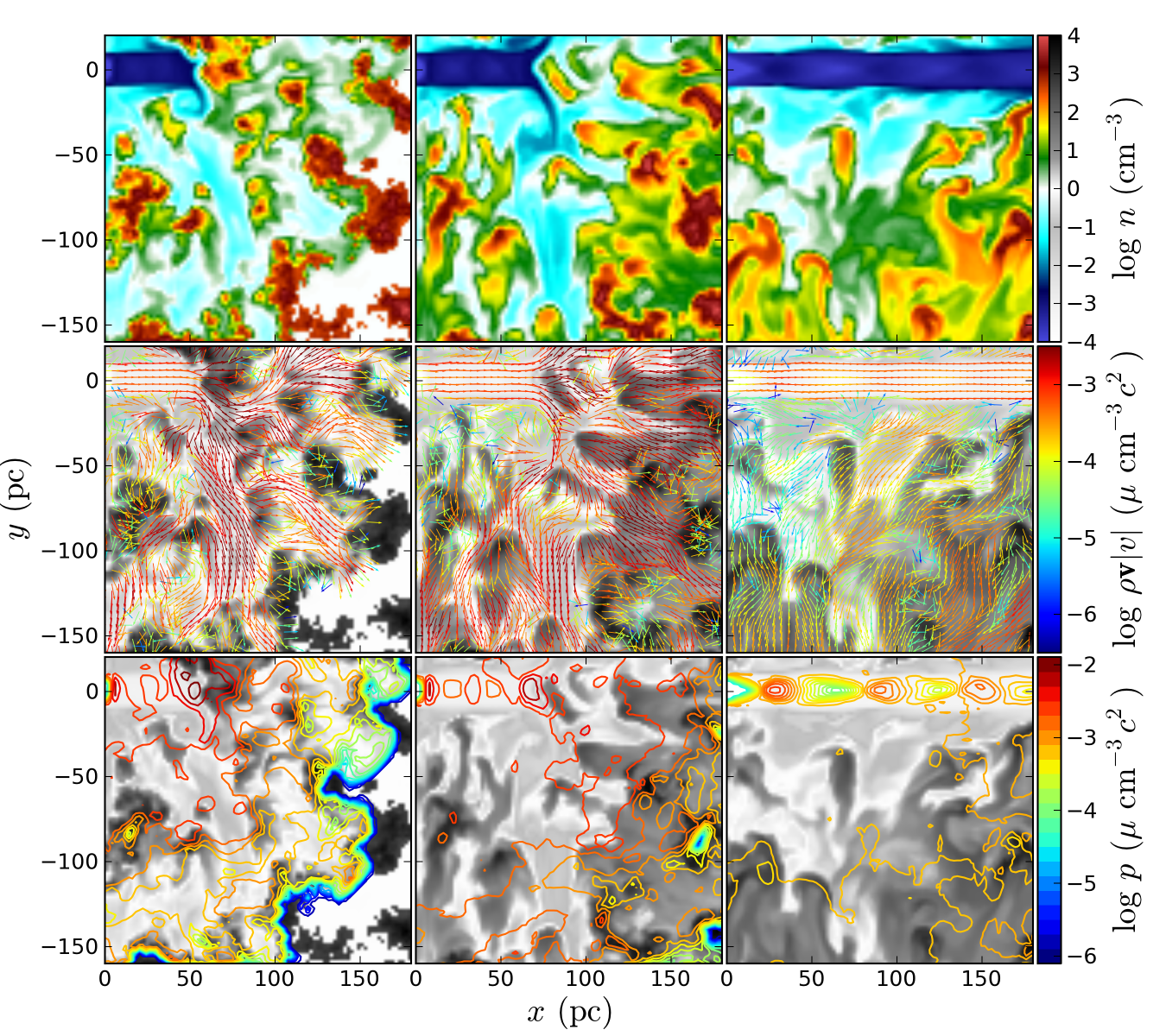}
  \end{center}
  \caption{Density and pressure maps of the mid-plane region near the jet base during three stages of jet-cloud interactions in run D$^\prime$. The upper row shows a color map of the density. In the middle and lower rows, ram pressure vectors $(\rho\mathbf{v}|{v}|)$ and thermal pressure contours $(p)$ are superimposed with a color scaling in units of $\mu \cmq c^2$ ($\mu$ is the mean mass per particle and $c$ is the speed of light) on a greyscale density map. The left, center, and right columns show the data at various times at which different effects dominate the acceleration mechanism of clouds: {\it Left}: the jet head is strongly interacting with the clouds in its path; {\it Middle}: hot-phase entraining jet streams carry ram pressure to clouds embedded in the bubble; {\it Right}: jet streams carrying entrained hot-phase and ablated cloud material dominate the channel flow. Compared to the ram pressure, the thermal pressure is relatively uniform inside the bubble. See the electronic edition of the Journal for a color version of this figure.}
  \label{f:nearjet}
\end{figure*}

Closer inspection of the ram pressure vectors ($\rho |\mathbf{u}| \mathbf{u}$), and thermal pressure gradients in the flow field in Fig \ref{f:nearjet} reveals a combination of effects that accelerate clouds faster than on a bubble evolution timescale. In the diverted secondary jet flow channels we do not see the large thermal pressure gradients maintained along the primary jet axis when the jet head encounters a cloud. It is evident from inspection of the density maps in Figs.~\ref{f:dens} and \ref{f:nearjet} that the mean density in the plasma flow that is flooding through the channels between the clouds is much larger than the mean density of a purely jet-blown bubble. Here the transfer of momentum is primarily maintained by a ram pressure that is comparable or somewhat greater to that of the primary jet flow by virtue of turbulent mixing (entrainment) of the shocked hot-phase gas with the jet plasma, and hydrodynamic ablation of cloud material into the engulfing flow. In channels with high mass-loading, the ram pressure even exceeds that of the primary jet. 

Another quantitative discrepancy with the theory contained in Eqn.~\eqref{t_acc} is that the flow velocities providing the ram pressure are not the expansion velocity of a fully thermalized bubble, $v_\mathrm{exp}$, but the velocity of partially thermalized channel flow, $v_\mathrm{ch}\sim10^5\kms$. 

Thus, two modifications need to be made to the above theory: 1) The density of the bubble in which the clouds are embedded is not solely determined by the mass injection of the jet, but the mass injection needs to be enhanced by the entrainment rate of hot phase material and the ablation rate of warm phase material; 2) The channel flow velocity that carries the momentum and provides the ram pressure at channel-cloud interfaces is not the expansion speed of the bubble but a much higher speed of only partially thermalized material. With these additions to the theory, we derive modified versions of Eqn.~\eqref{t_acc}.

In regard to point 2) above, we write $\vch$ instead of $\vexp$ on the RHS of Eqn.~\eqref{e:tacc1}:
\begin{equation}
\frac{\vexp}{\tacc} = \frac{3}{4} C_D \left( \frac{\rho_b}{\rho_c} \right) \frac{\vch^2}{R_c}\;. \label{e:taccch}
\end{equation}

If the channel flow in the bubble is entraining hot-phase material and mass-loaded by hydrostatic ablation of clouds, the mass injection into the bubble is augmented to 
\begin{equation}
\Mdottot=\Mdotjet+\Mdotentr+\Mdotabl\;,
\end{equation}
where the entrainment rate is approximately equal to the rate at which matter is swept up by the spherical bubble, 
\begin{equation}
\Mdotentr=4\pi\fentr\rho_a R_b^2\vexp\;,
\end{equation}
given the fraction of entrained material, $\fentr$. For the case of purely hydrostatic ablation\footnote{The ablation in this model is driven by pressure differences developing around the cloud surface, with ablation rates highest perpendicular to the flow direction.}, the ablation rate of one cloud with internal isothermal sound speed $c_c$ embedded in a channel flow with Mach number $\Mch$ \citep{hartquist1986b} is:
\begin{equation}
\mcdot = \alpha\min\left(1, \Mch^{4/3}\right)\left(m_c c_c\right)^{2/3}\left(\rho_b\vch\right)^{1/3}\;, \label{e:clabl}
\end{equation}
and the total mass injection rate into the bubble ablation rate is given by
\begin{equation}
\Mdotabl = \mcdot \times\frac{4\pi}{3}R_b^3\fvol\kmin^3\;.
\end{equation}
The constant $\alpha$ is of order unity for spherical clouds.

The treatment that follows is not entirely self-consistent because we do not determine the bubble expansion rate, $R_b(t)$, self-consistently under the modified conditions, but continue instead with the assumption that the outflow remains close to an energy-driven bubble that follows Eqn.~\eqref{e:rbub}. In \S\ref{s:bubble} we saw that this approximation is reasonable. The bubble density is obtained by integrating:
\begin{equation}
\td{}{t}\left(\frac{4\pi}{3}\rho_b R_b^3\right) = \Mdotjet+\Mdotentr+\Mdotabl \;,
\end{equation}
where $\Mdotjet$, $\Mdotentr$, and $\Mdotabl$ are defined by the equations above. This gives
\begin{equation}
\rho_b = \frac{3}{4\pi}A^{-3}\Mdotjet t^{-4/5} + \fentr\rho_a + \mcdot \fvol\kmin^3 \frac{5}{14} t \;. \label{e:rhob}
\end{equation}
Let us look at the case where $\Mdotentr \gg \Mdotjet$ and $\Mdotentr \gg \Mdotabl$, which holds early in the evolution of the bubble. We retain only the second term on the RHS in Eqn.~\eqref{e:rhob} and substitute that expression for $\rho_b$ into Eqn.~\eqref{e:taccch} to obtain:
\begin{equation}
\tacc = \frac{4}{3}C_D^{-1}\frac{\rho_c}{\fentr\rho_a}\vexp\frac{R_c}{\vch^2}\;.
\end{equation}
The acceleration timescale as a fraction of the dynamical time is then
\begin{equation}
\frac{\tacc}{t} = \frac{4}{5}C_D^{-1}\frac{\rho_c}{\fentr\rho_a}\frac{R_c}{\vch^2} A t^{-7/5} \;.\label{e:taccentr}
\end{equation}
Equation~(\ref{e:taccentr}) shows that clouds embedded in a heavily entrained spherical bubble will experience more efficient acceleration with time. Inserting typical values into Eqn.~\eqref{e:taccentr}, in particular, $\fentr=1$, and $\vch=\vchentr=10^5\kms$, we obtain $\tacc/t\sim0.078$ at $t=100\kyr$.

At $t=100\kyr$, the bubble density is, however, dominated by the mass loading from cloud ablation. We therefore turn to the limit $\Mdotabl \gg \Mdotjet$ and $\Mdotabl \gg \Mdotentr$. Retaining only the third term and using Eqn.~\eqref{e:clabl}, Eqn.~\eqref{e:rhob} becomes 
\begin{eqnarray}
\nonumber \rho_b &=& \alpha\min\left(1, \Mch^{4/3}\right)\left(m_c c_c\right)^{2/3}\left(\rho_b\vch\right)^{1/3} \fvol\kmin^3 \frac{5}{14} t \\
\nonumber &=& \frac{4\pi}{3} R_c^3 \rho_c c_c \min\left(1, \Mch^2\right) \vch^{1/2} \\
&&\times\left(\frac{5}{14} \alpha \fvol \kmin^3\right)^{3/2}t^{3/2} \;.\label{e:rhobabl}
\end{eqnarray}
We may estimate the channel speed from the (local) mass continuity condition that the mass ablation rate reach the channel mass flux through an area $\pi R^2$, $\mcdotmax = \rho_b \vchabl \pi R_c^2$, in which case, using $\rho_b = \mcdot \fvol\kmin^3\,(5/14)t$ (Eqn.~\ref{e:rhob}),
\begin{equation}
\vchabl = \frac{14}{5 \pi R_c^2 \fvol\kmin^3 } t^{-1} \;. \label{e:vchmin}
\end{equation}
Combining Eqns.~(\ref{e:taccch}), (\ref{e:rhobabl}), and (\ref{e:vchmin}), we obtain
\begin{eqnarray}
\nonumber \rho_b &=& \frac{10}{21} \left(\pi R_c^2\right)^{1/2} R_c \rho_c c_c \alpha^{3/2} \\
&&\times\min\left(1, \Mch^2\right) \fvol\kmin^3 t \;,\label{e:rhobabl2} \\
\nonumber \frac{\tacc}{t} &=& \frac{3}{14} C_D^{-1} \left(\pi R_c^2\right)^{3/2} c_c^{-1} \alpha^{-3/2} \\
&&\times\max\left(1, \Mch^{-2}\right) \left(\fvol \kmin^3\right)^{-1} A t^{-2/5} \:. \label{e:taccabl2}
\label{t_acc_mod}
\end{eqnarray}
Note that the expression for $\tacc/t$ (Eqn.~\ref{e:taccabl2}) only depends on the isothermal sound speed in the cloud, rather than directly on the density. In our simulations we observe that the channel flow dominated by entrained material moves with internal mach number, $\Mch\sim1$, while the channel flow dominated by ablated cloud material moves with internal Mach number $\Mch\sim3$. Taking $c_c=\sqrt{kT_c/\mu}$ with $T_c=100\Kv$, $\fvol=1$, $\muhot=0.6165$, and $\kmin=20\invkpc$, we find $\tacc/t=15.7$ for $\alpha=1$ and $\tacc/t=0.49$ for $\alpha=10$ at $t=100\kyr$.

\begin{figure*}
  \centering
  \figurenum{10}
  \includegraphics[width=0.43\textwidth]{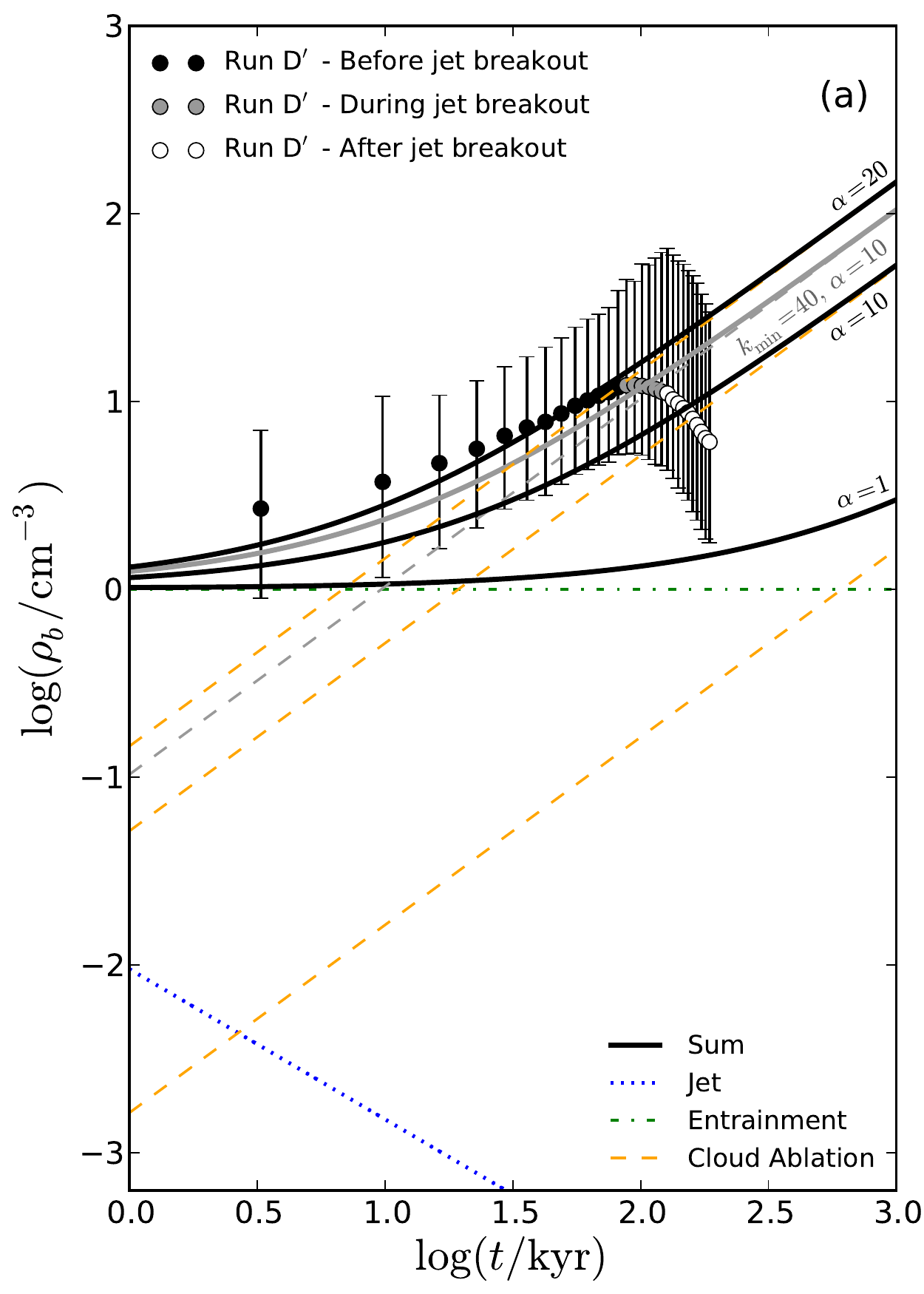}
  \includegraphics[width=0.43\textwidth]{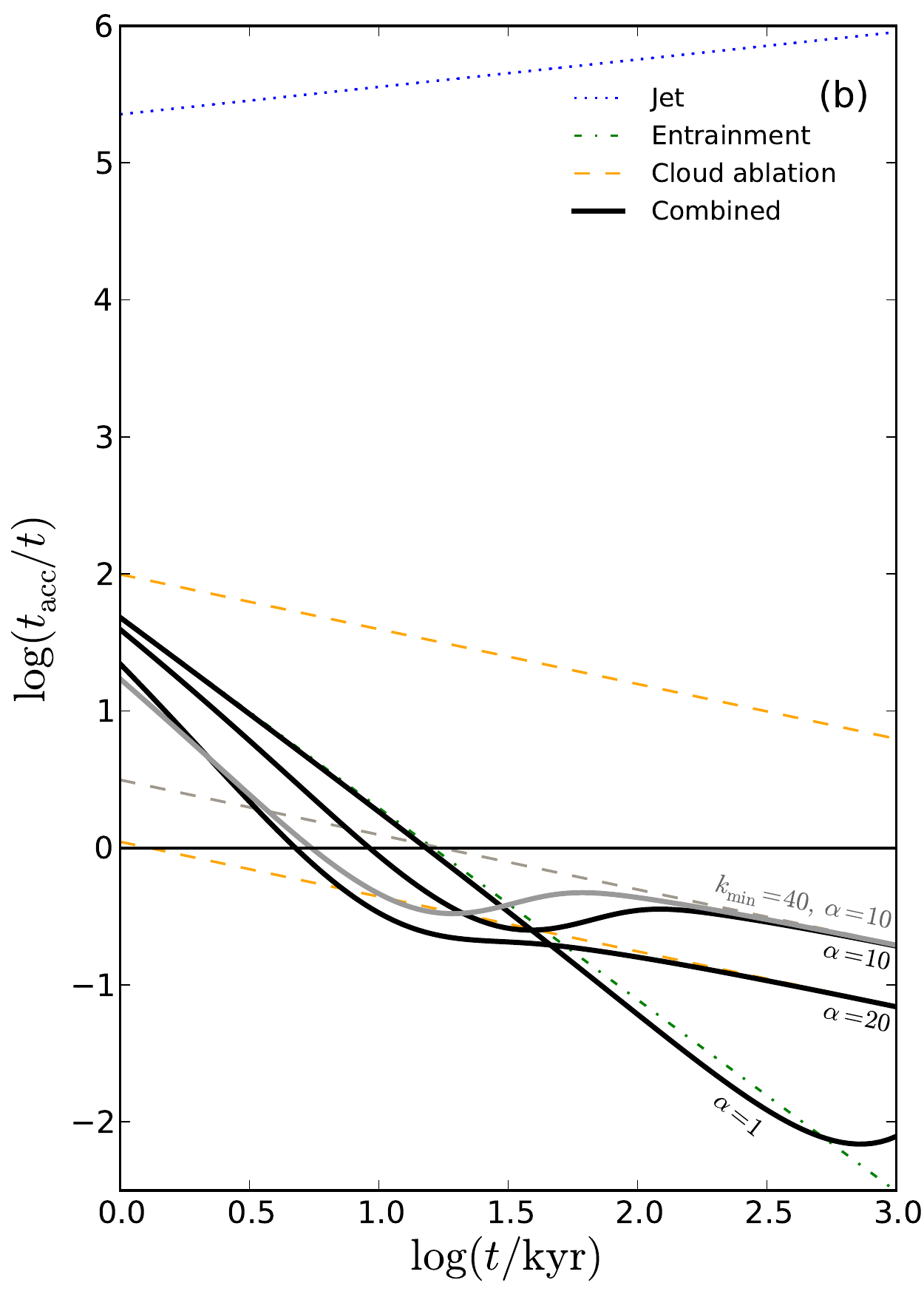}
  \caption{\textit{a}: The mean density of an energy-driven bubble as a function of time. Individual contributions to the density are shown in separate broken colored lines, as indicated in the legend. The combined contribution is shown in solid black lines for different values of $\alpha$ and $\kmin$. Superimposed are data points from Simulation D$^\prime$. The error bars denote estimate limits using different cutoff values for the tracer variable. \textit{b}: The ratio of cloud acceleration timescale to bubble age, versus time. The acceleration timescales as predicted for the cases that individual contributions dominate the bubble density and channel speed are indicated in broken colored lines. Note that the curves for $\alpha=10$ and for $\kmin=40\invkpc$, $\alpha=10$ for the cloud-ablation dominated case (dashed lines) overlap. The solid lines trace the (approximate) combined mean acceleration timescale, taking into account the transition from an entrainment-dominated bubble to a cloud-ablation dominated bubble. See the electronic edition of the Journal for a color version of this figure.}
  \label{f:tacc}
\end{figure*}

In Fig.~\ref{f:tacc}, we show the curves for the expressions for $\rho_b$ and $\tacc/t$ obtained above. Unless otherwise mentioned we used the set of typical parameters mentioned above. In the left panel we show the predicted mean bubble density for the case for which $\Mdotjet\gg\Mdotentr$ and $\Mdotjet\gg\Mdotabl$ (Eqn.~\ref{e:rhobjet} with Eqn.~\ref{e:mdotjet}) with a dotted blue line, the case for which $\Mdotentr\gg\Mdotjet$ and $\Mdotentr\gg\Mdotabl$, that is, $\rho_b=\fentr\rho_a$, with a dash-dotted green line, and the case for which $\Mdotabl\gg\Mdotentr$ and $\Mdotabl\gg\Mdotjet$ (Eqn.~\ref{e:rhobabl2}) with dashed orange lines. We see immediately that the maximum contribution to the total density by the jet plasma alone is much smaller compared to that of the other mass injection mechanisms. Since hot-phase entrainment is effective from $t=0$, jet plasma mass injection never dominates throughout the evolution of the bubble. Instead, the entrained material mixes well with the jet plasma, and the average bubble density is close to the ambient density. 

In time, cloud ablation becomes important in contributing to the total density. The bubble age at which this happens depends on $\rho_c$, $\kmin$, $R_c$, $\fvol$, and $\alpha$, and the thermodynamic warm phase parameters. We plot Eqn.~\eqref{e:rhobabl2} for $\alpha=1$, $\alpha=10$, $\alpha=20$, and for the case for which $\kmin=40\invkpc$, $R_c=12.5\pc$, $\alpha=10$. Superimposed are points from Simulation D$^\prime$, subdivided into the phases before, during, and after jet breakout. The error bars denote the upper and lower limits in the estimate of the mean bubble density in our simulations obtained by choosing different cutoffs for the tracer variables. The curves tracing the summed contributions to the total density are obtained by solving Eqn.~\eqref{e:rhob}, which is an implicit equation in $\rho_b$ and $t$, if all three terms on the RHS are included, with a standard root finding algorithm.

We find that cloud ablation is quite efficient in our simulations, requiring $\alpha>10$ to match the mean bubble densities seen in our simulations. \citet{hartquist1986b} introduced $\alpha$ as a constant near unity in the ideal case in which embedded clouds are spherical. Spatially fractal clouds have an (undefinably) larger surface area exposed to ablation, which leads to a larger inferred value for $\alpha$ from our simulations. The fact that the analytic curve for the case for which $\kmin=40\invkpc$, $R_c=12.5\pc$, $\alpha=10$ is similar to the curve for which $\kmin=20\invkpc$, $R_c=25\pc$, $\alpha=10$ lends support to the theory that the total surface area exposed to ablation by the channel flow is the main parameter governing the global bubble evolution, since a cloud distribution with $\kmin=40\invkpc$, $R_c=12.5\pc$ contains the same mass as a cloud distribution with $\kmin=20\invkpc$, $R_c=25\pc$ but four times the surface area for ablation.

In Fig.~\ref{f:tacc} b) we plot the acceleration timescale, $\tacc/t$ against $t$, for the same limiting cases as in Fig.~\ref{f:tacc} a). To evaluate the acceleration timescale when all mass injection terms are included, we use the solution to Eqn.~\eqref{e:rhob} and Eqn.~\eqref{e:vchmin} in Eqn.~\eqref{e:taccch}. Because of the strong dependence of the acceleration timescale on channel velocity (Eqn.~\ref{e:taccch}), we take into account that Eqn.~\eqref{e:vchmin} is only valid once cloud ablation becomes important by introducing an exponential turnover from $\vch=\vchentr$ to $\vch=\vchabl$ at $t=0$ over a timescale defined by equating the RHS of Eqn.~\eqref{e:taccabl2} with $\fentr\rho_a$. 

The acceleration timescale predicted by the model of clouds embedded in an expanding bubble (dotted lines) is too large to explain the cloud velocities seen in our simulations. The acceleration timescales predicted by ram-pressure driven, entrained and mass-loaded channel flows (solid lines), on the other hand, fall below $\tacc/t=1$ after $\sim10\kyr$ for all $\alpha$. In the early phases of the bubble evolution, the acceleration of clouds quickly becomes efficient as the hot-phase entraining jet plasma provides a high ram pressure to the clouds. The acceleration of clouds becomes somewhat less efficient when mass-loading from cloud ablation becomes important, as this reduces the channel flow velocity. 

The analytic estimates above only represent the global mean bubble density and mean cloud acceleration timescales in a spherically symmetric bubble, whereas, in reality, there exist radial and angular variations of the mean bubble density and channel speed, while clouds are driven outward. The relative contributions of jet plasma, hot-phase entrained channel flow, and warm-phase mass-loaded channel flow are, however, reasonably well explained by the analytic description when one allows for a high value of $\alpha$ due to the greater surface area of fractal clouds exposed to the ablating channel flow. Improvements to the analytic description may include a self-consistent evaluation of $R_b(t)$ and a finite spatial cloud distribution and finite cloud masses available for mass loading the channel flow, but such a treatment is beyond the scope of this paper.

We conclude that it is the ram pressure of mass-loaded channel flow, resulting from entrainment of the ambient medium and ablation of clouds, which provides the pressure gradients at cloud interfaces that transfer momentum and result in sustained high mechanical advantage and efficient bulk cloud acceleration within the dynamical time of the bubble. This acceleration mechanism of clouds found in our simulations may be interpreted as a variation of the two-stage feedback model proposed by \citet{hopkins2010a}. 

Hopkins \&{} Elvis proposed a radiative mode of quasar feedback, in which a radiation-driven quasar wind of the hot phase engulfs the warm and cold phase clouds in a galaxy and transfers energy and momentum to the clouds, accelerating and disrupting them through sustained pressure gradients and dynamical instabilities. The enlarged surface area of ablated and expanded clouds provide a larger cross-section for photoionization and radiation pressure from the quasar. Overall, this means that the efficiency of negative feedback is higher than that in the case of direct irradiation of clouds by the central AGN. The scenario the authors describe is very similar to the flow evolution and acceleration mechanisms we identified in our simulations. The jet plasma flow with its entrained hot phase material is analogous to the hot-phase quasar wind, and, as the flow engulfs and ablates clouds, sustained pressure gradients drive out the cloud material. In the case of our simulations, the pressure gradients are maintained by ram pressure, which may also be the case for quasar winds, in addition to radiation pressure.

\section{Discussion}\label{s:disc}

\subsection{The feedback efficiencies in radio galaxies with detected outflows}

\begin{deluxetable*}{lcccccccc}
\tablecaption{Radio galaxies with outflows\label{t:obs}}
\tablewidth{\textwidth}
\tablehead{
\colhead{Radio galaxy} & \colhead{$\log \, \Pjet$\tablenotemark{(a)}} & 
\colhead{$\vout$ (ionised) \tablenotemark{(b)}} & 
\colhead{$\vout$ (HI) \tablenotemark{(c)}} & 
\colhead{$z$\tablenotemark{(d)}} & \colhead{$\MBH$\tablenotemark{(e)}} & 
\colhead{Ref\tablenotemark{(f)}} & \colhead{$\eta$\tablenotemark{(g)}} \\
\colhead{} & \colhead{$(\erg)$} & \colhead{$(\kms)$} & \colhead{$(\kms)$} &  
\colhead{} & \colhead{$(\Msun)$} & \colhead{} & \colhead{} 
}
\startdata

\multicolumn{8}{c}{\citet{holt2008a}}\\
\hline
3C 213.1\dotfill    &   45.1$^\dagger$  &   142 & \nodata    & 0.19\phn & 9.1     & W09     & 0.0074       \\
3C 268.3\dotfill    &   45.9$^\dagger$  &   760 & \nodata    & 0.37\phn & 7.8     & W09     & 0.93\phn\phn \\
3C 277.1\dotfill    &   45.7$^\dagger$  &    79 & \nodata    & 0.320    & 7.6     & W09     & 0.89\phn\phn \\
4C 32.44\dotfill    &   46.5$^\ddagger$ &   852 & 128        & 0.368    & \nodata & \nodata & \nodata      \\
PKS 1345+12\dotfill &   44.6$^\dagger$  &  1980 & 400        & 0.122    & 7.8     & W09     & 0.052\phn    \\
3C 303.1\dotfill    &   45.5$^\dagger$  &   438 & \nodata    & 0.27\phn & 8.4     & W09     & 0.11\phn\phn \\
PKS 1549-79\dotfill &   45.3$^\dagger$  &   679 &  79        & 0.152    & 8.0     & W09     & 0.15\phn\phn \\
PKS 1814-63\dotfill &   45.6$^\ddagger$ &   162 &  24$^\ast$ & 0.065    & 8.7     & M11     & 0.058\phn    \\
PKS 1934-63\dotfill &   45.6$^\ddagger$ &    93 & \nodata    & 0.182    & \nodata & \nodata & \nodata      \\
PKS 2135-20\dotfill &   46.8$^\ddagger$ &   157 & \nodata    & 0.636    & \nodata & \nodata & \nodata      \\
PKS 2314+03\dotfill &   45.8$^\dagger$  &   497 & 350        & 0.220    & 8.5     & W09     & 0.15\phn\phn \\
\hline\noalign{\smallskip}
\multicolumn{8}{c}{\citet{nesvadba2006a,nesvadba2008a,nesvadba2010a}$^{1,2,3}$}\\
\hline
MRC 1138-262$^1$\dotfill & 46.2        &  800              & \nodata & 2.16 & 8.7 & N06      & 0.16\phn \\
MRC 0316-257$^2$\dotfill & 46.0        &  670              & \nodata & 3.13 & 8.3 & S07      & 0.40\phn \\
MRC 0406-244$^2$\dotfill & 46.2        &  960              & \nodata & 2.44 & 8.5 & S07$^\S$ & 0.40\phn \\
TXS 0828+193$^2$\dotfill & 46.4        &  800              & \nodata & 2.57 & 8.7 & S07$^\S$ & 0.40\phn \\
3C 326$^3$\dotfill       & 45\phd\phn  & 1800$^{\ast\ast}$ & \nodata & 0.09 & 8.6 & HR04     & 0.020    \\
\hline\noalign{\smallskip}
\multicolumn{8}{c}{\citet{torresi2012a}}\\
\hline
3C 445\dotfill   & 44.9 & 10000$^{DW}$             & \nodata & 0.0562 & 8.3 & M04  & 0.028\phn \\
3C 390.3\dotfill & 45.1 & \phn\phn600\phm{$^{DW}$} & \nodata & 0.0561 & 8.6 & WU02 & 0.029\phn \\
3C 390.3\dotfill & 45.1 & 43769$^{DW}$       & \nodata & 0.0561 & 8.6 & WU02 & 0.029\phn \\
3C 382\dotfill   & 44.8 & \phn1000\phm{$^{DW}$}    & \nodata & 0.0579 & 9.1 & M04  & 0.0044    \\
\hline\noalign{\smallskip}
\multicolumn{8}{c}{\citet{guillard2012a}}\\
\hline
3C 236\dotfill      & 45.9 &  507    & 750 & 0.10\phn & \nodata & \nodata & \nodata \\
3C 293\dotfill      & 45.7 &  494    & 500 & 0.045    & 8.0     & W09     & 0.40    \\
3C 305\dotfill      & 45.8 & \nodata & 250 & 0.042    & 7.9     & WU02    & 0.63    \\
3C 459\dotfill      & 46.4 &  372    & 300 & 0.23\phn & 8.5     & W09     & 0.63    \\
4C 12.50\dotfill    & 45.4 &  812    & 600 & 0.12\phn & 7.8     & W09     & 0.31    \\
IC 5063\dotfill     & 45.3 & \nodata & 350 & 0.011    & 7.4     & V10     & 0.63    \\
OQ 208\dotfill      & 43.7 & \nodata & 600 & 0.077    & \nodata & \nodata & \nodata \\
PKS 1549-79\dotfill & 45.9 &  906    & 250 & 0.15\phn & 8.0     & W09     & 0.63    \\
\hline\noalign{\smallskip}
\multicolumn{8}{c}{\citet{stockton2007a}}\\
\hline
3C 48\dotfill   & 46.5$^\dagger$ & 491  & \nodata & 0.369 & 8.8 & W09  & 0.39\phn

\enddata
\tablenotetext{(a)}{Jet power. $^\dagger$Values from \citet[][Table 1]{wu2009b}. $^\ddagger$Jet powers neither given in reference or in \citet{wu2009b} were computed using the $1.4\GHz$ flux and the scaling relation by \citet{cavagnolo2010a}.}
\tablenotetext{(b)}{Outflow velocity of ionized gas. $^{\ast\ast}$Terminal velocity estimated from Na D absorption blueshift of $350\kms$. $^{DW}$Disk wind.}
\tablenotetext{(c)}{Outflow velocity of neutral gas, seen in \HI{} absorption. $^\ast$Revised by \citet{morganti2011a}.}
\tablenotetext{(d)}{Redshift.}
\tablenotetext{(e)}{SMBH mass.}
\tablenotetext{(f)}{Reference or calculation method for SMBH mass. W09: SMBH mass taken from \citet[][Table 1]{wu2009b}; S07, HR04: SMBH mass estimated from stellar bulge mass, $\Mbulge$, given in \citet[][S07]{seymour2007a} or \citet[][HR04]{haring2004a}, respectively, using the Magorrian relation $\MBH=0.0015\Mbulge$. $^\S$$\Mbulge$ is only an upper limit;  M04: \citet{marchesini2004a}; WU02: \citet{woo2002b}; V10: \citet{vasudevan2010a}; M11: \citet{morganti2011a}; N06: using the stellar mass estimated by \citet{nesvadba2006a} and the Magorrian relation, $\MBH=0.0015\Mbulge$.}
\tablenotetext{(g)}{Jet Eddington ratio, $\eta=\Pjet/\Ledd$ estimated using $\MBH$.}
\end{deluxetable*}

There are a growing number of observational studies of radio galaxies in which outflows of cold neutral or warm ionized material are detected. We have compiled a set of 27 radio galaxies from the samples studied by \citet{holt2008a}, \citet{nesvadba2006a,nesvadba2008a, nesvadba2010a}, \citet{torresi2012a}, \citet{guillard2012a}, and the individual case of 3C 48 by \citet{stockton2007a}. We also use the sample of \citet[][data obtained through private communication]{lehnert2011a} in the following comparison with our simulation. The galaxies are listed in Table~\ref{t:obs} whose columns include the jet power, the outflow velocities in either neutral or ionized gas (or both), the SMBH mass, and the value of $\eta=\Pjet/\Ledd$ estimated from the SMBH mass. 

When jet powers were not given in the references above, they were either found in the table of properties (Table 1) of the sample compiled by \citet{wu2009b} or calculated from the $1.4\GHz$ flux available from the NASA Extragalactic Database using the scaling relation by \citet{cavagnolo2010a}.

We note that the none of the estimates for any of these quantities are very accurate. The jet power is typically only an order of magnitude estimate. While outflow velocities are usually accurate to a few tens of percent, the velocities only represent a particular phase of gas moving at the given bulk velocity. The accuracy of the black hole mass estimates depend on the method used but, for this sample, they are reliable to within a factor of a few. Given these uncertainties, it is not possible to arrive at strong conclusions, but the method of comparison itself is of some interest.

\citet{holt2008a} studied the optical emission line gas kinematics in 14 compact radio sources and found strong evidence for fast outflows in 11 cases. 8 of the 11 cases are also known to have blueshifted \HI{} absorption. We list both neutral and ionized outflow velocities in Table~\ref{t:obs}. The two fastest outflows at $\sim2000\kms$ and $\sim850\kms$ were found in GPS sources, and the authors report a general trend in their sample that the larger outflow velocities were seen in more compact objects, although orientation also plays a role.

\citet{nesvadba2006a,nesvadba2008a} observed four HzRGs ($z\sim2$ -- $3$) with the SINFONI integral spectroscopy unit on the VLT and detected red- and blueshifted \OIII{} emission aligned with the jet axis. These galaxies contain large ionized gas masses (few $10^{10}\Msun$) comparable to the mass of the neutral and molecular gas. The inferred outflow speeds of ionized gas were $\sim600$ -- $1000\kms$ and energy coupling efficiencies between the jet and outflow were of order 0.1. \citet{nesvadba2010a} posit that star formation in the comparatively nearby radio galaxy 3C 326 is maintained low by the energy input of the radio jet, which drives out a fast outflow of ionized gas and keeps the neutral and molecular gas warm, in what is sometimes termed the \lq\lq{}maintenance phase\rq\rq{} of AGN feedback. HzRGs, on the other hand, may be in an \lq\lq{}establishment phase\rq\rq{} of galaxy formation, in which stellar populations are born in starbursts that are then abruptly shut down by the AGN jet. 

\citet{torresi2012a} describe the properties of the only three broad line radio galaxies in which outflows of typically $100$ to $1000\kms$ have been detected through absorption lines in the soft X-ray spectrum. These AGN are said to contain \lq\lq{}warm absorbers\rq\rq{} \citep{blustin2005a, mckernan2007a}. They find two sources harboring outflows originating in the torus or narrow line region rather than being associated with a disc wind. Their data extend the tentative positive correlation for Type 1 QSOs between radio loudness and warm absorber mass outflow rate. We include these two sources in our sample in consideration of the possibility that the acceleration of the outflows is driven by jet-ISM interactions.

Warm absorbers are observationally distinct from the ultra-fast outflows (UFOs) detected in radio quiet AGN at high incidence \citep[at least 35\%,][]{tombesi2010b} through highly ionized and blueshifted Fe K-shell absorption lines in the hard X-ray band \citep{cappi2006a}. UFOs are thought to be mildly relativistic disc winds ($\sim0.01c$ to $0.1c$, several $10^3\kms$ to $10^4\kms$) with mass outflow rates comparable to the accretion rate, and outflow kinetic luminosities a significant fraction of the bolometric luminosity \citep{tombesi2012a}. The discovery of UFOs in radio-loud AGN \citep{tombesi2010b,tombesi2011a} blurs the distinction between disc winds and jets and motivates research into unified theories of disc-wind-jet systems. 3C 390.3, for example, harbors a UFO with outflow velocity $(0.146\pm0.004)c$, in addition to the warm absorber. Studies of jet/wind-disc interactions have focused more on galactic microquasars \citep{fender2004a,neilsen2009a}, often regarded as scaled down versions of AGN \citep{mirabel1999a}, and many simulation models are applicable to both microquasars and AGN \citep{takeuchi2010a,ohsuga2011a}. UFOs have kinetic luminosities comparable to those of AGN jets, so they may have a significant impact on the galaxy-scale ISM of the host. They tend to be much less collimated and somewhat slower than jets at the same radius, but the physics of the interactions with the ISM may be similar to that presented in this study. This needs to be verified with commensurate simulations of ultra-fast disc winds expanding from sub-parsec to galactic scales in an inhomogeneous two-phase ISM.

\citet{guillard2012a} have detected outflows in ionized gas through broad blueshifted [\NeII] emission lines in Spitzer IRS observations of a sample of radio galaxies, for which previous studies found neutral outflows in \HI{} absorption. We list both neutral and ionized outflow velocities in Table~\ref{t:obs}. Five of the sources have highly blueshifted wings on the [\NeII] line that match with the blueshifted broad \HI{} absorption.

\citet{lehnert2011a} fitted the NaD absorption feature of 691 SDSS sources with extended radio morphologies, redshift $z<0.2$, and $1.4\GHz$ fluxes greater than $40\mJy$, finding modestly blueshifted but highly broadened absorption excesses for about half the sources. The authors deduce the presence of outflows with a distribution of terminal velocities between $150\kms$ -- $1000\kms$, mean mass and energy outflow rates of $10\Msun\yr^{-1}$ and $10^{42}\ergs$, respectively.

\citet{stockton2007a} mapped the \OIII{} emission near the central region of 3C 48, a powerful $z\approx0.369$ CSS source and ULIRG born from a major merger \citep{scharwachter2004a}, using the GMOS integral field unit on Gemini North. The \OIII{} emission is blueshifted by $\sim500\kms$ and offset by $\sim1\kpc$ northward of the quasar, along the jet axis. It is, therefore, distinct from the AGN narrow line region. The stellar age estimate of the host by \citet{stockton2007a} disfavors the hypothesis that the current jet activity triggered the starburst, but the energetics and the alignment with the jet of the outflowing gas support the view that a substantial amount of material is driven out by the jet as it is breaking out of the dense central region. 

Two sources feature in both the samples of \citet{holt2008a} and \citet{guillard2012a}, PKS 1345+12 (4C 12.50) and PKS 1549-79. The estimates for the outflow velocities are different in each study so we list them here separately. In addition to the outflow observed in the neutral phase through \HI{} absorption \citep{morganti2004a} and in the ionized phase \citep{spoon2009b}, the ULIRG and GPS source 4C 12.50 exhibits outflows of $\sim600$ -- $1000\kms$ in warm and cold molecular gas \citep{dasyra2011a, dasyra2012a}. The observations for the molecular outflows are not included in the table but the data suggest that 4C 12.50 may be a remarkable case of a radio AGN in which the all phases of the ISM are strongly coupled to the relativistic jet. For sources for which a black hole mass was found in the literature we calculated the value of $\eta=\Pjet/\Ledd$.

\begin{figure}
  \figurenum{11}
  \includegraphics[width=\linewidth]{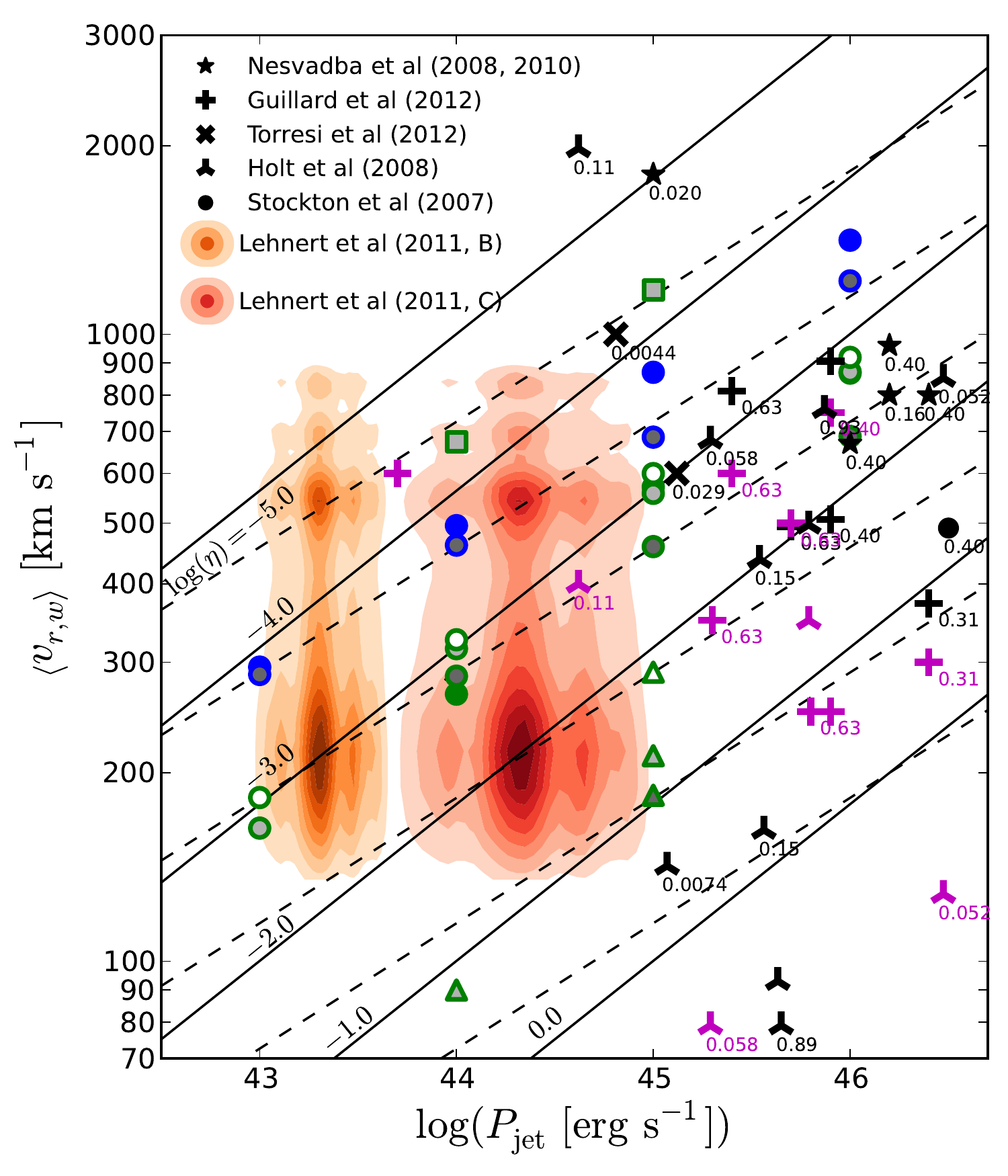}
  \caption{Simulation results and data from observations compiled in Table \ref{t:obs} on the plane defined by the outflow speed and jet power. Simulation results are shown in blue and green points corresponding to ISM hot phase densities $\nhot=0.1\cmq$ and $1.0\cmq$, respectively. Triangle, round, and square markers denote $\kmin=10\invkpc$, $20\invkpc$, and $40\invkpc$, respectively, while the shading of the marker from filled, through dark and light grey, to white denote filling factors of 0.42, 0.13, 0.052, and 0.027, respectively. The samples studied by the various authors are marked with different symbols and contours as shown in the legend. The black symbols mark the measured outflow speeds of ionized material, and the magenta symbols mark the measured outflow speeds of neutral material. Subscripts denote the value of $\eta$. Lines representing the \msigma{} relation for constant values of $\eta$ are superimposed. The solid and dashed lines for which the power and intercepts of the \msigma{} relation $\log_{10}(\MBH/\Msun) = B + \delta\log_{10}(\sigma/200\kms)$ are $(\delta, B) = (4.0, 8.1)$, $(5.0, 8.1)$, respectively. See the electronic edition of the Journal for a color version of this figure.}
  \label{f:obs}
\end{figure}

We present the data from the compiled observations together with those from our simulations on the plane defined by outflow velocity and jet power in Fig.~\ref{f:obs}. The data by \citet{lehnert2011a} are plotted as filled contours in probability density, $\ud N / \ud \Pjet \ud \vout$. The orange contours show the probability density if the scaling between $1.4\GHz$ flux and jet power by \citet{birzan2004a} is used \citep[see also][]{best2006a}, and the red contours show the probability density if the scaling by \citet{cavagnolo2010a} is used. The data of the galaxies compiled in Table \ref{t:obs} are shown with different symbols as indicated in the legend. Black markers denote outflow velocity measurements for ionized gas, while magenta markers indicate velocities for outflows of neutral gas. The points are labeled with the value for $\eta=\Pjet/\Ledd$, for the cases for which the mass of the SMBH was found in the literature. Lines of constant $\eta$ for \msigma{} relations $\log_{10}(\MBH/\Msun) = B + \delta\log_{10}(\sigma/200\kms)$, for which the power and intercepts are $(\delta,B) = (4.0, 8.1)$, $(5.0, 8.1)$, are superimposed in solid and dashed, respectively. Simulation points are shown in blue and green with different shades of fill color denoting the different filling factors: $\fvol=0.42$, 0.13, 0.052, 0.027 correspond to filled, dark grey, light grey, and white. Square, round, and triangular markers represent simulations for which $\kmin=40\invkpc$, $20\invkpc$, $10\invkpc$, respectively.

The interpretation of the location of a point for the outflow of an observed galaxy on this plane is the same as that for the simulation points. If the point lies above a given line of constant $\eta$, feedback is effective in that galaxy if that value $\eta$ applied to that galaxy. For some galaxies we have calculated the value of $\eta=\etaobs$, so a direct comparison between that and the critical value $\etacrit$ marked by its location with respect to the lines of constant $\eta$ can be made. We find that most galaxies with outflows listed in Table \ref{t:obs} lie below the line $\etacrit>10^{-3}$. For most galaxies $\etaobs>\etacrit$, meaning that feedback associated with the outflows is effective in these galaxies.

\begin{figure}
  \figurenum{12}
  \includegraphics[width=\linewidth]{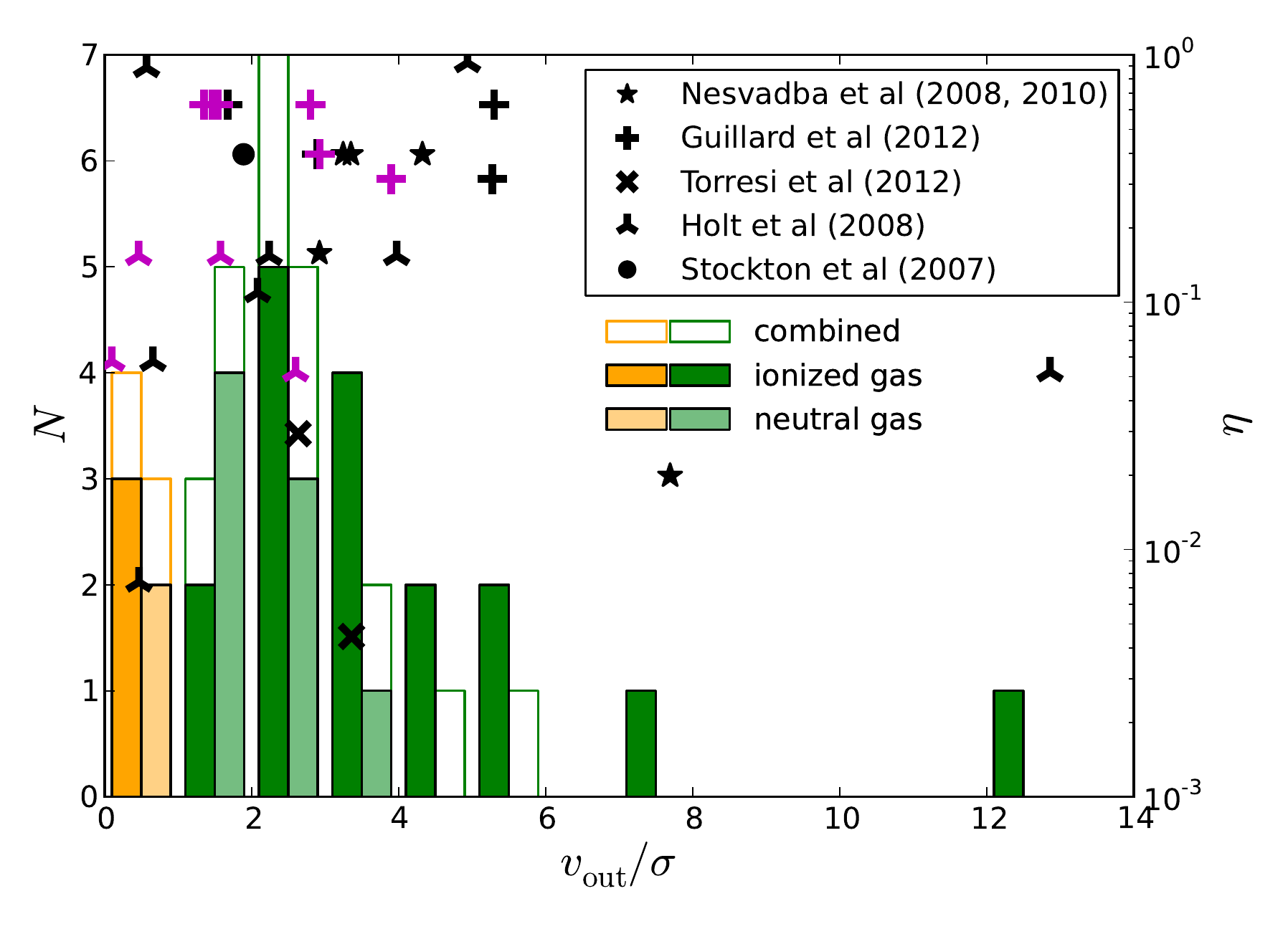}
  \caption{The absolute frequency $N$ of the ratio of outflow velocity to the velocity dispersion, $\vout/\sigma$,  as predicted by the \msigma{} relation for all radio galaxies in Table \ref{t:obs}, for which the outflow velocity and SMBH mass are available. The 14 bins in $\vout/\sigma$ each have a width of unity. The first orange bin counts the number of sources for which feedback is not effective, while the green bins count the sources in which feedback is effective. The darker colored bars denote outflow velocities of ionized gas and the lighter colored bars those measured for the neutral gas component in the galaxies. The unfilled bars show the combined distribution for all outflow velocity measurements; those above the darker and lighter bars, are, respectively, for the distributions for which the measurement of the neutral gas was preferentially take over that for the ionized gas, and vice versa. Overlaid are the values of $\eta$ (right ordinate axis) for each binned galaxy. See the electronic edition of the Journal for a color version of this figure.}
  \label{f:veta}
\end{figure}

The effectiveness of feedback is better seen in terms of the distribution of the ratio of observed outflow velocity to velocity dispersion predicted by the \msigma{} relation for the galaxies, $\vout/\sigma$. Here we assume an \msigma{} relation for which the power and intercept are 5.3 and 8.2, respectively, values that give the best fit to elliptical galaxies, according to \citet{graham2011a}. The distribution is shown as a histogram Fig.~\ref{f:veta}. Each of the 14 bins has a width of unity. Thus, the bin in orange to the left of $\vout/\sigma=1$ counts the number of sources for which feedback is not effective (according to the criterion used in this paper), whereas the green bins to the right of $\vout/\sigma=1$ count the sources for which feedback is effective. Again we differentiate between observations of the ionized and neutral phases; the left, darker bars in a bin count the number for measurements of the outflow velocities in ionized phase, and the right, lighter bars count those for the neutral phase. The unfilled bars trace the combined distribution of outflow velocities. The unfilled bars above the bars for the ionized (neutral) gas outflow velocity distribution show the combined distribution for which the measurement of the ionized (neutral) outflow velocity was preferentially taken over the measurement of the neutral (ionized) outflow velocity. Scattered on the same plot against the right ordinate axis are values for $\eta$ for each of the binned items.

From Fig.~\ref{f:veta} we find that the outflows in most galaxies have velocities that are a factor of a few greater than a velocity dispersion predicted by the \msigma{} relation. Only in a few galaxies the outflows are weak in relation to the mass of the galaxy. There is no correlation between the values of $\eta$ for galaxies in which feedback is effective and in those in which feedback is not effective. In both domains, the scatter is large. However, most galaxies have large values of $\eta$, with $\eta>10^{-2}$. The locations of the points for the observed galaxies in Fig.~\ref{f:obs} in relation to those of our simulations suggest that the ISM distribution in radio galaxies where fast outflows are observed is clumpy on quite large scales ($\Rcmax>25\pc$) and possibly quite dense with a mean hot phase density of $1\cmq$ within the inner $1\kpc$.

Because of the small sample size and large uncertainties associated with observationally measured jet powers, outflow velocities, and SMBH masses, none of the conclusions above are definitive. The comparison between observations and simulations presented here are merely a first step at understanding how effective feedback is in radio galaxies at different redshift. A benefit of this method is that one may draw conclusions about the gas distribution of the hosts' ISM without direct spatially resolved observations. Conversely, if spatially resolved observations of the hosts' ISM become possible with, e.g. the Atacama Large Millimeter/submillimeter Array \citep[ALMA,][]{planesas2011a,lonsdale2012a}, or the Square Kilometer Array \citep[SKA,][]{godfrey2012a}, the parameter space of relevant simulations and the derived feedback efficiencies can be strongly constrained. A similar comparative approach as demonstrated here between theory and observations could be made for radio quiet AGN that feature other modes of feedback.

\subsection{Other negative and positive feedback criteria}\label{s:othf}

The criterion for effective negative feedback adopted in this work that the warm-phase outflow velocities driven by AGN jets exceed the velocity dispersion predicted by the \msigma{} relation for a galaxy is not unique. The chosen criterion, however, has the advantage that it is directly relevant to the evolution of the ISM in the bulge of galaxies \citep{kormendy2009b}, and that simulation and observation may be superimposed on the same plane spanned by velocity and jet power.  One alternative is to use the condition that the material in the jet-driven outflow reaches the escape velocity and will be unbound from the potential of a galaxy. The fate of the outflowing gas is an interesting question in itself, and will need to be determined by simulations on scales of $10$ -- $1000\kpc$ simulations. Furthermore, the fulfilment of either condition does not necessarily imply inhibited star-formation. 

Positive feedback is more difficult to quantify because the conditions for star formation are governed by a range of competing physics influencing the gravitational stability of clouds and the formation of dense cold cores within. They include external pressurization, hydrodynamically and conductively driven ablation, shocks driven into clouds, X-ray ionization, and molecular chemistry, and non-ideal MHD effects. Jet-induced star-formation is thought to occur in gas-rich galaxies and proto-clusters at high redshifts \citep{miley2008a}, but there are also examples in the nearby universe, e.g. in Centaurus A \citep{mould2000a}, and Minkowski's object \citep{croft2006a}, and 3C 285 \citep{van-breugel1993a}, in which the radio jet is implicated in shock- and pressure-triggered star formation. 

\citet{gaibler2011a} showed that the star-formation rate in a disc galaxy is enhanced a factor of 2 -- 3 as a result of jet-ISM interactions, primarily because the pressure bubble driven by the jet compresses the gas in the disc. We also observe an increase in the probability distribution of dense gas in our simulations. Both external compression and (under-resolved) radiative shocks contribute, and to properly assess whether the a cloud would collapse to form stars, it is necessary to perform simulations with self-gravity. WB11 compared the mass of the clouds to the critical Bonnor-Ebert mass before and after the clouds were engulfed by the high-pressure jet plasma and concluded that the external pressurization places the clouds in the unstable regime, but also that the cloud ablation time-scales were short compared to the collapse timescales. A similar conclusion was reached by \cite{antonuccio-delogu2008a} with simulations of a jet passing near an isolated cloud. The fact that \citet{gaibler2011a} observe a marked increase in the star formation rate in the compact disk of molecular gas with high filling factor is consistent with the picture that cloud ablation is less important the larger the cloud complexes are, as described in \S\ref{s:fvol} and \S\ref{s:kmin}. The simulations by \cite{sutherland2007a} also demonstrate that dense gas distributed in a disc-like geometry couples less readily to the jet in the form of an outflow. It is clear that the consequences of AGN jet feedback (and other modes of feedback) depend as much on the multi-phase ISM properties of the galaxy during feedback as on the power of the central source.

\subsection{Cloud ablation}

The ablation and destruction of clouds is a difficult process to capture accurately in hydrodynamical simulations. There have been many studies investigating the destruction of clouds overrun by shocks (e.g. in supernova remnants), or clouds embedded in a flow (e.g. a stellar wind) with numerical simulations, taking into account a variety of physical effects including cooling \citep{vietri1997a,cooper2009a,yirak2010a}, thermal conduction \citep{orlando2005a,orlando2008a}, structural inhomogeneity \citep{xu1995a,poludnenko2002a,nakamura2006a,cooper2009a}, and magnetic fields \citep[][]{gregori1999a,shin2008a}. Rayleigh-Taylor and Kelvin-Helmholtz instabilities create a turbulent interface between the cloud surface and ambient flow where the mixing between the two phases occurs, but resolution limitations and the artificial viscosity due to numerical diffusion in hydrodynamic simulations clamp the spatial scales and energy scales that can be captured, thereby diluting the mixing process \citep{pittard2009a}. 

Since the comprehensive analytic and numerical work by \citet{klein1990a,klein1994a}, the fiducial minimum resolution to capture the complete destruction of an adiabatic spherical uniform cloud in hydrodynamical simulations has been accepted to be around 100 cells. Radiative cooling, however, radically changes the destruction mechanism of the clouds \citep{vietri1997a,cooper2009a}. In most astrophysical flows including those in our simulations, the cooling timescale is much shorter than the flow crossing time, and radiative shocks driven into clouds develop a thin, protective wall near the edge of the cloud boundary, which inhibits the Kelvin-Helmholtz instability from rapidly destroying the cloud. \citet{cooper2009a} showed that the cloud breaks up into long-lived filamentary cloudlets advected far downstream. 

%

While individual clouds in our simulations are under-resolved, it is not clear by how much we are systematically overestimating or underestimating cloud ablation. The fact that we obtain a value of the ablation coefficient defined by \citep{hartquist1986b}, $\alpha$, in Eqn.~\eqref{e:clabl} of order 10, rather than 1, from the predicted mean bubble density at late times of our simulations may indicate that we are overestimating the cloud ablation rate in our simulations, perhaps because the stabilizing cooling layers behind the clouds are insufficiently resolved. 

The development of turbulence and the degree of fragmentation during shock-cloud and shock-wind interactions depend on the physical structure of the clouds \citep{xu1995a,poludnenko2002a,nakamura2006a,cooper2009a}. \citet{nakamura2006a}, for example, find that cloud destruction is prolonged for smoother interfaces between clouds and the embedded medium. On the other hand, \citet{cooper2009a} observe that fractal clouds fragment faster than spherical clouds. The large value of ablation coefficient may, therefore, also be a realistic consequence of the inhomogeneous, fractal outlines of our clouds, which seed the Kelvin-Helmholtz instabilities and initially enhance the ablation rate. The ablation and entrainment rate seen in our simulations may also be higher because the flow, in which the clouds are embedded, is already rather turbulent as the jet streams percolate through the inter-cloud channels. Pressure variations and \lq\lq{}buffeting\rq\rq, particularly in the wake of the cloud may be efficient at extracting material from the back of the clouds and entraining it into the tail streams. This effect was observed by \citet{pittard2009a} to lead to faster cloud destruction.

\citet{pittard2010a} directly compare the lifetimes of clouds in their simulations with those predicted by Eqn.~\eqref{e:clabl} and find that, for clouds with low density contrasts embedded in low Mach number flows the expression underestimates the ablation rate by a factor of $\sim4$, while for clouds with high density contrasts embedded in high Mach number flows the predicted values are a factor of $\sim2$ -- $5$ larger than found in the simulations. This may also partially account for the large value of $\alpha$ found in this work.

The statistical distribution of clouds in our simulations for which $\kmin=20\invkpc$, is identical to that used by \citet{sutherland2007a}, who performed a three level resolution study showing that the fractal features in their simulations were at least partly captured. As demonstrated by \citet{stone1992a} and \citet{klein1994a} for spherical adiabatic clouds, \citet{cooper2009a} found in their study of individual radiative fractal clouds embedded in a supersonic flow that the rate of fragmentation depended on the resolution of the cloud. However, no convergence was found at the highest resolutions. A recent resolution convergence study by \citet{yirak2010a}, probing a broader range of resolutions of radiative shock-cloud interactions, found that, unlike adiabatic cases, the flow structure does not show signs of convergence at 100 cells per clump; convergence may only formally be reached when the cooling length is well resolved. This is highly impractical in global simulations such as those conducted in this work. One possible improvement discussed by \citet{yirak2010a}, although complicated, involves a careful use of adaptive mesh refinement of the cooling layers in the radiative shocks. Another possibility is to implement a subgrid treatment of turbulence that leads to better convergence with increased resolution \citep{pittard2009a}.

Given the difficulty in ensuring sufficient resolution to reach flow convergence and the complexity of including all the physical effects that may modify the cloud destruction timescale in opposing ways, the setup of our simulations, despite limited resolution across one cloud are justified as a first step to model AGN jet feedback in fully three-dimensional hydrodynamic simulations.

\subsection{Long-term evolution}

The domain extent in our simulations covered the central $1\kpc^3$ of a gas rich radio galaxy, within which the coupling of radio jet to the dense gas is strongest. The fate of the outflowing gas on scales larger than $10\kpc$ will influence the accretion and star-formation rates of the galaxy at later times and is also relevant to theories of the enrichment of the intracluster medium (ICM). An example in which an AGN radio jet may be directly responsible for carrying enriched material into the ICM was identified in 4C+44.16 by \citet{hlavacek-larrondo2011a}. But the fraction of gas that is unbound from the galaxy cannot be predicted from our simulations and separate simulations on larger scales with the inclusion of a gravitational field need to be performed. In this problem, non-uniformity of the dense gas distribution and asymmetries in the energy injection influence the requirements to unbind gas from a deep gravitational potential \citep{bland-hawthorn2011a}. An inhomogeneous ISM and off-center energy injection reduce the fraction of gas that can reach escape velocities. 

In a gravitating environment, buoyantly rising jet-inflated bubbles and associated buoyancy-driven instabilities \citep{balbus1989a} will influence the mixing rate and evolution of the ISM \citep{bruggen2002a}. Large-scale simulations will also aid comparisons with observations of the radio jet and the outflowing gas, which have limited resolution on kpc scales. 

The balance between heating and cooling in the ICM of cool-core clusters can also be tested. For example, \citet{gaspari2012a} performed grid-based hydrodynamic simulations on these scale and found that a feedback cycle involving a heavy, slow jet reproduced the typical observed entropy profiles around cool core cluster central galaxies. Such heavy, slow jets may be the result of entrainment and mass-loading that jets experience in the core of galaxies though interactions similar to those shown in our simulations. Gaspari et al. also demonstrated the pervasiveness of a two-phase ISM owing to cold phase gas condensing out of the hot phase through the thermal instability and fuelling the AGN.

\section{Conclusions}\label{s:concl}

We have conducted a total of 29 3D grid-based hydrodynamical simulations of AGN jets interacting with a fractal two-phase ISM. The simulations cover jet powers of $\Pjet=10^{43}\ergs$ -- $10^{46}\ergs$ and a range of different ISM parameters characterized by the density of the hot phase, the filling factor of the warm phase (clouds), and the maximum cloud sizes in the fractal distribution of the warm phase. The simulations are applicable to the early and intermediate phases of radio-mode feedback at high redshifts, which often involve massive, evolving gas-rich (proto)-galaxies. The sufficiently broadly sampled parameter space allowed us to study the dependence of the negative feedback efficiency, as measured by the maximum outflow velocities reached through the jet-ISM interactions, on the filling factor and the maximum clouds sizes. We also identified the precise physics of the momentum and energy coupling that leads to rapid acceleration of the warm phase. Finally, we undertook a comparison between recent observations of outflows in radio galaxies and our simulations on the plane defined by outflow velocity and jet power.

Fifteen new simulations were conducted to supplement the simulation data by WB11 with results for feedback efficiencies for filling factors of 0.052 and 0.027, and those with maximum cloud sizes of $\Rcmax=10\pc$ and $\Rcmax=50\pc$. The lower filling factor runs and runs for which the maximum cloud sizes were $\Rcmax=10\pc$ were conducted at a resolution of $1\pc$ per cell, twice the resolution of the simulations by WB11 within the same box size. The principal findings from these new simulations are the following:
\begin{enumerate}
  \item The main conclusions reached in WB11 remain unchanged: Feedback is effective in systems in which the jet power is in the range $\Pjet=10^{43}$ -- $10^{46}\ergs$ and $\eta>\etacrit$, where $\eta$ is the Eddington ratio of the jet, $\eta=\Pjet/\Ledd$. The critical Eddington ratio, $\etacrit$, is determined by the criterion that the velocity of the outflow driven by a jet of power $\Pjet$ exceed the velocity dispersion expected from the \msigma{} relation of a galaxy for which the Eddington luminosity is $\Ledd$. For reasonable values of the ISM parameters, feedback ceases to be efficient for Eddington ratios of $\eta\lesssim10^{-4}$. Due to a large sustained mechanical advantage in these systems, the fraction of jet energy transferred to the warm phase is $\sim0.1$ -- $0.4$.
  \item The dependence of the feedback efficiency on filling factor found by WB11 reverses for filling factors $\fvol\lesssim0.1$ in the sense that a smaller filling factor leads to higher feedback efficiencies. In general the dependence on filling factor is weak, however. The reversal occurs because of a shift in the balance of two opposing effects as the filling factor is decreased: the increase in surface area of the clouds exposed to ablation relative to their volume, and the increase in the volume between clouds through which the jet plasma may flood. The latter decreases the feedback efficiency while the former increases the feedback efficiency and dominates below $\fvol\sim0.1$.  
  \item For a given filling factor, the feedback efficiency is higher the smaller the maximum cloud sizes in the fractal distribution. Galaxies containing cloud complexes initially larger than $50\pc$ ($\kmin=10\invkpc$) require jets with Eddington ratios $\eta>10^{-2}$ for efficient negative feedback. Pressure triggered star-formation may be expected for large clouds ($\gtrsim50\pc$), while clouds smaller than $\sim10\pc$ ($\kmin=40\invkpc$) are unlikely to survive significant ablation. The dependence of feedback efficiency on cloud size is much stronger than that on filling factor and scales nearly linearly with $\kmin$. This is the result of the linear relation between the size of clouds and the surface area of the clouds exposed to ablation relative to their volume. We introduce the concept of an interaction depth for jet-cloud interactions analogous to optical depth. The interaction depth increases linearly with smaller cloud sizes leading to a linear increase in outflow velocity with $\kmin$.
  \item A comparison between the global dynamics of the outflowing warm phase material with that of an energy-driven bubble shows that outflows approach the energy-driven limit in cases for which the filling factor is low or the maximum sizes of clouds is large. For a given filling factor, the dispersal of clouds is higher if clouds are smaller. Conversely, a jet-driven bubble impacting and engulfing a distribution of large clouds may lead to pressure-triggered star-formation. Thus, the size distribution of clouds strongly influences whether feedback is negative or positive. Considering the relatively weak dependence of bubble expansion rate on cloud sizes, we argue that, if conditions in the ISM of radio galaxies in cosmological simulations are in the regime of $\fvol\lesssim0.1$ and $\Rcmax\lesssim25\pc$, an energy-driven sub-grid implementation of (negative) AGN radio mode feedback is justified. In this regime, feedback in a single phase medium is a good approximation to feedback in a two-phase medium because the warm phase material embedded in the energy-driven bubble is accelerated to speeds comparable to the bubble expansion speed within the dynamical time of the bubble.
  \item We find that a simple theory of clouds embedded in a fully thermalized energy-driven bubble does not provide sufficient ram pressure to accelerate the clouds to velocities observed in high redshift radio galaxies. Instead, the momentum transfer is provided by the ram-pressure of the partially thermalized streams of jet plasma flooding through the inter-cloud channels, which have turbulently entrained ambient hot gas (initially external to the expanding bubble) and are mass-loaded with ablated warm cloud material. Initially the channel flows carry turbulently entrained shocked hot-phase material at velocities of $\sim10^5\kms$, and at later times, the channel flows are dominated by material ablated from clouds. The acceleration efficiency is highest in the former stage, while mass-loading from clouds reduces the acceleration efficiency somewhat. This mechanism, which is reminiscent of the two-stage feedback mechanism proposed by \citet{hopkins2010a}, is capable of accelerating the clouds to velocities of $100$s to $1000$s $\kms$ within the dynamical time of the bubble.
  \item The observed outflows of neutral and ionized material in most of the radio galaxies in our sample compiled from the literature are fast enough to cause substantial velocity dispersions in the host. The critical jet Eddington ratios in most sources is $\etacrit\gtrsim10^{-3}$, but the observed jet Eddington ratios are predominantly $\eta\gtrsim10^{-2}$. By comparing the jet powers and outflow speeds obtained from the observations with those found in our simulations, we tentatively infer that the ISM of the radio galaxies in which outflows are observed is clumpy on quite large scales ($\Rcmax>25\pc$) and possibly quite dense with a mean (hot phase) density of $1\cmq$ within the inner $1\kpc$.
  \item We explain some of the radio-morphological characteristics of 3C 48 discussed by \citet{stockton2007a} with a synthetic radio image of one of our simulations, including the jet collimation within the extended \OIII{} emission region, and the deflection and expansion of the jet lobe beyond the emission region.
\end{enumerate}



\acknowledgments
We thank the referee for a prompt, thorough, and constructive report, which greatly improved the clarity of the paper. The computations for this paper were undertaken  on the NCI National Facility at the Australian National University. The software used in this work was in part developed by the DOE-supported ASC / Alliance Center for Astrophysical Thermonuclear Flashes at the University of Chicago. We are grateful to Dr. Ralph Sutherland who provided revised cooling functions for this work and code to generate fractal cubes for the initial warm-phase distribution. We thank Ajay Limaye (NCI) for creating the volume rendered visualizations in Fig.~\ref{f:vol}. We thank Matthew Lehnert for providing us with the data used in the NaD absorption study of nearby radio galaxies \citep{lehnert2011a}. This research has made use of the NASA/IPAC extragalactic database (NED).  During the course of this research AYW was in part supported by a Japanese Society for the Promotion of Science (JSPS) fellowship (PE 11025). This work was supported in part by the {\it FIRST} project based on Grants-in-Aid for Specially Promoted Research by MEXT (16002003) and JSPS Grant-in-Aid for Scientific Research (S) (20224002).

\bibliographystyle{apj}

\end{document}